\documentclass[longauth]{aa}  

\usepackage{graphicx}
\usepackage{xcolor}
\usepackage[fleqn]{amsmath}
\usepackage{txfonts}
\begin{document}

   \title{Double, double, toil, and trouble:}

   \subtitle{The tails, bubbles, and knots of the local compact obscured nucleus galaxy NGC\,4418}

   \author{C.\,F.\,Wethers\inst{1} \thanks{\email{wethers@chalmers.se}},
          S. Aalto\inst{1},
          G.\,C.\,Privon\inst{2,3,4},
          F.\,Stanley\inst{5},
          J.\,Gallagher\inst{6},
          M.\,Gorski\inst{1},
          S.\,König\inst{1},
          K.\,Onishi\inst{1},
          M.\,Sato\inst{1},
          C.\,Yang\inst{1},
          R.\,Beswick\inst{7},
          L. Barcos-Munoz\inst{8}
          F.\,Combes\inst{9},
          T.\,Diaz-Santos\inst{10},
          A.\,S.\,Evans\inst{11,8},
          I.\,Garcia-Bernete\inst{12},
          C.\,Henkel\inst{13,14},
          M.\,Imanishi\inst{15},
          S.\,Martín\inst{16,17},
          S.\,Muller\inst{1},
          Y.\,Nishimura\inst{18},
          C.\,Ricci\inst{19,20},
          D.\,Rigopoulou\inst{2},
          \and
          S.\,Viti\inst{21,22};\\
          }

   \institute{Department of Space, Earth and Environment, Chalmers University of Technology, Onsala Space Observatory, 439 92 Onsala, Sweden
        \and National Radio Astronomy Observatory, Charlottesville, VA 22903, USA 
        \and Department of Astronomy, University of Florida, Gainesville, FL 32611, USA 
        \and Department of Astronomy, University of Virginia, Charlottesville, VA 22904, USA 
        \and The Institut de radioastronomie millimétrique (IRAM), Domaine Universitaire, 38406 Saint Martin d’Hères, France 
        \and Wisconsin IceCube Particle Astrophysics Center, Madison, WI 53703, USA
        \and Jodrell Bank Centre for Astrophysics, Department of Physics and Astronomy, The University of Manchester, M13 9PL, UK
        \and Department of Astronomy, University of Virginia, 530 McCormick Road, Charlottesville, VA 22903, USA
        \and Observatoire de Paris, LERMA, Collège de France, CNRS, PSL University, Sorbonne University, Paris, France
        \and Institute of Astrophysics, Foundation for Research and Technology Hellas (FORTH), Heraklion, GR-70013, Greece
        \and National Radio Astronomy Observatory, 520 Edgemont Road, Charlottesville, VA 22903, USA
        \and Department of Physics, University of Oxford, Keble Road, Oxford OX1 3RH, UK
        \and Max-Planck-Institut für Radioastronomie, Auf dem Hügel 69, 53121 Bonn, Germany
        \and Astronomy Department, Faculty of Science, King Abdulaziz University, P. O. Box 80203, Jeddah 21589, Saudi Arabia
        \and National Astronomical Observatory of Japan, National Institutes of Natural Sciences (NINS), 2-21-1 Osawa, Mitaka, Tokyo 181-8588, Japan
        \and European Southern Observatory, Alonso de Córdova, 3107, Vitacura, Santiago, 763-0355, Chile
        \and Joint ALMA Observatory, Alonso de Córdova, 3107, Vitacura, Santiago, 763-0355, Chile
        \and Department of Astronomy, The University of Tokyo, 7-3-1, Hongo, Bunkyo, Tokyo 113-0033, Japan
        \and Nu\'{c}leo de Astronom\'{i}a de la Facultad de Ingenier\'{i}a, Universidad Diego Portales, Av. Ej\'{e}rcito Libertador 441, Santiago 22, Chile
        \and Kavli Institute for Astronomy and Astrophysics, Peking University, Beijing 100871, People’s Republic of China
        \and Leiden Observatory, Leiden University, PO Box 9513, 2300 RA Leiden, The Nether- lands
        \and Department of Physics and Astronomy, University College London, Gower Street, WC1E 6BT, London, UK  
             }
   \date{Received ????; accepted ????}

%$^1$Department of Space, Earth and Environment, Chalmers University of Technology, \\ 
%            Onsala Space Observatory, 439 92 Onsala, Sweden \\ 
 %           email: {\tt wethers@chalmers.se} \\[\affilskip]
            
%            $^2$National Radio Astronomy Observatory, 520 Edgemont Rd., Charlottesville, VA 22903, USA  \\[\affilskip]
            
%            $^2$Department of Astronomy, University of Florida, P.O. Box 112055, Gainesville, FL 32611, USA  \\[\affilskip]
            
%            $^2$Department of Astronomy, University of Virginia, 530 McCormick Rd., Charlottesville, VA 22904, USA  \\[\affilskip]
            
%            $^3$Institut de Radioastronomie Millimétrique, 300 rue de la Piscine, Domaine Universitaire \\
%            38406 Saint Martin d’Hères, France \\[\affilskip]
            
%            $^4$Wisconsin IceCube Particle Astrophysics Center,
%            222 West Washington Ave., Suite 500, Madison, WI 53703, USA}
% \abstract{}{}{}{}{} 
% 5 {} token are mandatory
 
  \abstract
   {Compact obscured nuclei (CONs) are an extremely obscured (N$_{\rm{H2}} >$10$^{25}$ cm$^{-2}$) class of galaxy nuclei thought to exist in 20-40 per cent of nearby (ultra-)luminous infrared galaxies While they have been proposed to represent a key phase of the active galactic nucleus (AGN) feedback cycle, the nature of these CONs -- what powers them, their dynamics, and their impact on the host galaxy -- remains unknown.}
   %This work analyses the large-scale optical properties of the local CON, NGC\,4418 ($z$=0.00727). We explore the impact of nuclear outflows on the host galaxy, map the spectral features of NGC\,4418 for the first time and infer the power source of the CON in its centre. We aim to discover new structures in the galaxy and determine the powersource of the CON. For the first time, we aim to map the spectral features of the galaxy.
   {This work analyses the galaxy-scale optical properties of the local CON NGC\,4418 ($z$=0.00727). The key aims of the study are to understand the impact of nuclear outflows on the host galaxy and infer the power source of its CON. Through the mapping of the galaxy spectra and kinematics, we seek to identify new structures in NGC\,4418 to ultimately reveal more about the CON's history, its impact on the host, and, more generally, the role CONs play in galaxy evolution.}
   {We present new, targeted integral field unit observations of the galaxy with the Multi-Unit Spectroscopic Explorer (MUSE). For the first time, we mapped the ionised and neutral gas components of the galaxy, along with their dynamical structure, to reveal several previously unknown features of the galaxy.}
   {We confirm the presence of a previously postulated, blueshifted outflow along the minor axis of NGC\,4418. We find this outflow to be decelerating and, for the first time, show it to extend in both directions from the nucleus. We report the discovery of two further outflow structures: a redshifted southern outflow connected to a tail of ionised gas surrounding the galaxy and a blueshifted bubble to the north. In addition to these features, we find the [O\,{\sc iii}] emission reveals the presence of knots across the galaxy, which are consistent with regions of the galaxy that have been photoionised by an AGN.}
   {We identify several new features in NGC\,4418, including a bubble structure, a reddened outflow, and [O\,{\sc iii}] knot structures throughout the galaxy. We additionally confirm the presence of a bilateral blueshifted outflow along the minor axis. Based on the properties of these features, we conclude that the CON in NGC\,4418 is most likely powered by AGN activity.}


    \titlerunning{NGC\,4418 with MUSE}
    \authorrunning{C. F. Wethers et al.}
   \maketitle
%
%***************************************************************************
\section{Introduction}
\label{sec:introduction}
%***************************************************************************

Compact obscured nuclei (CONs) have been shown to exist in as many as 40 per cent of nearby (ultra-)luminous infrared galaxies \citep[(U)LIRGs; e.g.][]{falstad21,garcia22}. These compact (<100 pc) nuclear structures are characterised by their extreme nuclear column densities, N$_{\rm{H2}}$ $>$10$^{25}$ cm$^{-2}$ \citep[e.g.][]{sakamoto13}, which make them almost invisible at mid-IR, optical, and even X-ray wavelengths due to the heavy attenuation \citep[e.g.][]{treister10,lusso13,roche15}. Currently, the only way to identify CONs is through the rare $\nu$2 = 1f transition of vibrationally excited HCN (HCN-VIB), which is usually radiatively excited by intense mid-IR emission from hot dust at high column densities \citep{aalto15b}. While a handful of alternative identification methods are being developed \citep[e.g.][]{donnan23}, including using the 1 millimetre surface brightness \citep{falstad21} and the silicate absorption slope \citep{garcia22}, no CONs have yet been confirmed via these methods.

Although the exact nature of CONs remains unknown, they likely play a critical role in the evolution of galaxies as an early, obscured starburst phase in which remnant material is being expelled from the centre of the galaxy via, for example, active galactic nucleus (AGN) driven outflows. Such a phase could represent the early stages of a feedback cycle, where a fraction of the expelled gas will eventually fall back into the galactic plane. Indeed, while only a handful of CONs have been confirmed in the nearby Universe  (z$<$0.05), all show signatures of molecular outflows in HCN and/or CO, and several host molecular gas inflows \citep{falstad21}. Even in the CONs with no obvious inflow signatures, the CO outflows appear to be slow (V$<$400 km\,s$^{-1}$), meaning this gas will likely remain within the galaxy. This inflow and outflow feedback cycle provides a key mechanism by which a super-massive black hole (SMBH) affects its host galaxy on scales larger than its sphere of influence. Thus, identifying these CON outflows, understanding what drives them, and quantifying their effect on the larger-scale galaxy properties is critical in building a co-evolutionary model of galaxies.

Two main interpretations of the CON phenomenon exist. The first is a {hidden AGN} scenario, in which the nuclear region is powered by active accretion onto the central SMBH. In this scenario, the galaxy nucleus is completely obscured by remnant material, following a merger or interaction event \citep[e.g.][]{kocevski15, ricci17, blecha18, boettcher20}. The frequency of such interaction events, and their severity, has been shown to increase with (U)LIRG luminosity \citep{sanders96}, meaning that if CONs are associated with rapid gas inflow onto a SMBH, they should reside in more luminous LIRGs and merger-rich environments. Indeed, \cite{falstad21} find the CON fraction of ULIRGs (L$_{\rm{IR}}>$ 10$^{12}$\,L$_{\odot}$; $\sim$40 per cent) is almost double that of LIRGs (10$^{11}< $L$_{\rm{IR}}>$10$^{12}$\,L$_{\odot}$; $\sim$20 per cent). Furthermore, the compact nature of CONs strongly favours a hidden AGN scenario \citep[e.g.][]{evans03,costagliola10,sakamoto10}. The second interpretation is that CONs are powered by a nuclear nascent starburst, deeply embedded in a dusty cocoon. This so-called {embedded starburst} scenario was initially put forward to explain the unusual far-IR and radio properties observed in CONs \citep[e.g.][]{yun01,bressan02,roussel03}. The scenario postulates that CONs host a very young starburst ($\sim$4-6 Myr) with little to no non-thermal radio emission arising from supernovae (SNe). These starbursts are rare by definition, owing to their short lifespans, and become 'ordinary' starbursts containing many SNe. However, several groups have argued that dust is unlikely to completely cover an extended starburst to such an extent as to cause the deep absorption features observed in CONs \citep{roche86, dudley97, spoon01}. 

Distinguishing between the embedded starburst and hidden AGN scenario for CONs has proven challenging, in part due to the fact that the extreme nuclear column densities of CONs can completely erase many of the traditional AGN signatures in the galaxy spectra. Nevertheless, identifying the dominant power source of CONs remains critical in building a cohesive model of galaxy evolution. AGN are currently considered a short, but reoccurring, phase in the lifetime of a galaxy. However, if CONs are found to be universally powered by AGN, this would increase the predicted number of local AGN by a factor of $\sim$2 \citep{maiolino03} and thus revolutionise our understanding of AGN phase(s) in the evolution of local galaxies. Alternatively, if CONs are powered by a nascent starburst, they represent a rare and unusually intense compact nuclear starburst, which likely plays an important role in the growth of galactic nuclei. 

NGC\,4418 ($z$=0.00727) has been described as a {prototypical CON} \citep[e.g.][]{costagliola13} and has been the subject of several studies into the nature of CONs. NGC\,4418 is part of an interacting system with the nearby dwarf galaxy VV655, which lies to the south-east at a distance of $\sim$30 kpc (2-3 arcmin) from NGC\,4418 \citep[e.g.][]{varenius17,boettcher20} and may be in part responsible for several of the outflowing and inflowing features observed throughout the galaxy. A study by \cite{sakamoto13} finds evidence for an outflow on scales of $\sim$1 kpc along the minor axis of NGC\,4418. More recently, \cite{ohyama19} confirm the presence of this outflow and classify it as a {dusty superwind}, based on both shock-heated emission lines and the enhanced stellar NaD absorption. Additionally, work by \cite{gonzalez12b} finds a substantial inflow in NGC\,4418 with a mass inflow rate $\dot{M}$ $\gtrsim$ 12 M$_{\odot}$yr$^{-1}$. The existence of both inflows and outflows in NGC\,4418 implies that its nucleus is evolving very rapidly, yet the structure of these features and their impact on the galaxy remain unknown. As a result, identifying the mechanism(s) by which cool gas is funnelled into the centre of the NGC\,4418 has been the subject of several recent studies. Within the local galaxy population, such gas infall events are likely triggered by major gas-rich mergers, yet to date observations of NGC\,4418 have shown none of the prominent tidal tails, shells, or stellar asymmetries typically associated with such interactions. Instead, \cite{boettcher20} postulate the trajectory of the neighbouring dwarf galaxy, VV655, has likely caused tidal torques in NGC\,4418, leading to material from the initial gas disk being driven towards the galaxy centre. Indeed, models demonstrate that a satellite galaxy with a Small Magellanic Cloud-like mass, such as VV655, can induce torques sufficient to trigger a significant inward migration of gas from the disk \citep{pettitt16,ramon20}. Confirming the mechanism(s) by which so much dust and gas migrates to the galaxy centre remains an important question, both for determining the nature of CONs and galaxy evolution in general.

Despite extensive multi-wavelength studies, the nuclear power source of NGC\,4418 is also unknown. On the one hand, X-ray observations with \textit{Chandra} show NGC\,4418 to have a flat, hard X-ray spectrum, usually indicative of a Compton-thick AGN \citep{maiolino03}. This scenario is supported by the apparent compactness of the CON in NGC\,4418, with near- and mid-IR observations indicating an optically thick source with a diameter $<$70 pc \citep{evans03}, and vibrationally excited HC$_3$N \citep{costagliola10} and HCN \citep{sakamoto10} studies deriving sizes $<$10 pc in diameter. On the other hand, \cite{roussel03} suggest the low radio-to-far-IR luminosity ratio of NGC\,4418 is entirely consistent with a young starburst ($<$5 Myr),  and \cite{varenius14} find several compact nuclear structures believed to be super star clusters. Many of these structures, however, lie outside the optically thick 5\,$\mu$m photosphere identified from the near-IR CO absorption from warm molecular gas \citep{ohyama23}, within which most of the galaxy luminosity is expected to arise. It is therefore unlikely that the SNe and super star clusters seen by \cite{varenius14} are the dominant CON power source. With the advent of optical integral field unit (IFU) observations with optical instruments such as the Multi-Unit Spectroscopic Explorer (MUSE), we may be able to distinguish between these scenarios in NGC\,4418 for the first time. Despite the heavy obscuration at their nuclei, optical IFU data allow  the kinematics of the ionised and neutral gas to be spatially resolved and shed light on the inflow and outflow structures and their impact on the host galaxy. 

Here we present the first results from new IFU observations of NGC\,4418 with MUSE. In section~\ref{sec:data} we outline the observations, reduction pipeline, and data products. Section~\ref{sec:results} presents our key findings, including the 2D mapping of the dust extinction, dynamical structure, and inflow and outflow features in NGC\,4418. Possible interpretations of these results are discussed in section~\ref{sec:discussion} and are summarised in section~\ref{sec:conclusions}. Throughout this paper, we assume a flat $\Lambda$ cold dark matter cosmology with $H_{0}$ = 70 km\,s$^{-1}$ Mpc$^{-1}$, $\Omega_{M}$ = 0.3, and $\Omega_{\Lambda}$ = 0.7. We adopt a distance of 34 Mpc, at which 1 arcsec = 165pc \citep{sakamoto13}. Unless otherwise specified, all quoted magnitudes are given in the AB system.

%***************************************************************************
\section{Data}
\label{sec:data}
%***************************************************************************

    %-----------------------------------------------------------------------
    \subsection{Data overview}
    \label{sec:data_overview}
    %-----------------------------------------------------------------------

    We present new observations from MUSE for the local CON galaxy, NGC\,4418. MUSE is a panoramic IFU spectrograph on the Very Large Telescope (VLT), based at Cerro Paranal, Chile. The IFU capabilities of MUSE allow the kinematics of the ionised and neutral gas to be mapped throughout the galaxy for the first time. Observations of NGC\,4418 were acquired in 2020 February, during ESO period 104 (ESO0104.B-0668; P.I. F. Stanley). Observations were carried out in wide-field mode (WFM), covering a 60 $\times$ 60 arcsec field of view (FoV) with a spatial sampling of 0.2 $\times$ 0.2 arcsec and a theoretical instrument spatial resolution of 0.4 arcsec (full-width at half-maximum; FWHM) at 7000\,\r{A}. The nominal instrument setup provides a spectral coverage from 4750 to 9300\,\r{A} at a mean resolution of 2.5\,\r{A} (FWHM) and provides a dataset of $>$90,000 individual spectra (323 $\times$ 324 spaxels). The total on-source integration time was 2.8\,h, with an average seeing of 0.8 arcsec. The raw data products were reduced via the MUSE \textsc{EsoReflex} pipeline (v.2.6.2) to produce a fully calibrated and combined science-ready data cube with a point spread function (PSF) FWHM of 0.8$\times$1.0 arcsec at 5000\,\r{A}. Any residual sky contamination was identified using the Zurich Atmosphere Purge (ZAP) software package (Soto et al. 2016) and subtracted from the final data cube.
    
    %-----------------------------------------------------------------------
    \subsection{Continuum subtraction}
    \label{sec:continuum_subtraction}
    %-----------------------------------------------------------------------

    Following the initial reduction of the MUSE data cube, we fitted and subtracted the stellar contribution from the galaxy continuum in order to isolate light from the emission features. The continuum fitting makes use of single stellar population (SSP) synthesis models from v9.1 of the Medium resolution INT Library of Empirical Spectra \citep[MILES;][]{falcon11}. The complete MILES library consists of spectra for $\sim$1000 stars spanning a large range in atmospheric parameters, obtained with the 2.5m INT telescope. The available stellar templates span ages 0.03 - 17.78 Gyr and metallicities 0.0001 $<Z<$ 0.04. The SSP templates provide composite spectral energy distributions (SEDs) comprised from multiple stellar population templates, each with a single age and metallicity. Each template spans wavelengths 3525-7500\,\r{A} \citep{sanchez06} with a spectral resolution of 2.5 \r{A} (FWHM), corresponding to the mean instrumental resolution of MUSE. Templates are fitted pixel-by-pixel to the MUSE data cube using a penalised pixel fitting (pPXF) method, based on a non-linear least squares fitting routine \citep{cappellari17}. As part of this process, the spectral data were re-binned such that each channel had a fixed velocity width, $\Delta$V = 55.9 km\,s$^{-1}$. Throughout the fitting, all strong emission lines in the MUSE spectra are masked to prevent them from biasing the fit of the stellar templates and overestimating the continuum level. In this way, we ensured the fitted stellar continuum templates are constrained only by the strength of the stellar absorption and the galaxy continuum, which traces the local interstellar reddening, and the age and metallicity of the stellar populations. 

    Following the continuum modelling and subtraction there remains a non-negligible H$\beta$ ($\lambda$4861$\AA$) absorption feature in several spaxels. This residual absorption is centrally concentrated. It appears strong towards the galaxy nucleus, but becomes negligible at larger radii. To account for this residual absorption in the continuum-subtracted data cube, we performed a spaxel-by-spaxel non-linear least squares fitting over the spectral region 4750-4950 \r{A}, covering the H$\beta$ feature. The model for the fitting combines two Gaussian functions: a broad, negative component to model the H$\beta$ absorption, and a narrower positive component to model the H$\beta$ emission. During the fitting, we accounted only for spaxels in which the residual H$\beta$ absorption is significant, namely, $>$3$\sigma$ of the spectral noise as measured from a nearby, featureless region of the spectrum. The absorption component of the best-fit model is then removed from the original continuum-subtracted cube to produce an H$\beta$-corrected data cube. Unless otherwise stated, all analysis in this work is performed on the continuum-subtracted data cube corrected for any residual H$\beta$ absorption.

%***************************************************************************
\section{Results}
\label{sec:results}
%***************************************************************************
    %-----------------------------------------------------------------------
    \subsection{Overview}
    \label{sec:results_overview}
    %-----------------------------------------------------------------------

\begin{figure}
    \centering 
        \includegraphics[trim= 15 160 15 200,clip,width=.5\textwidth]{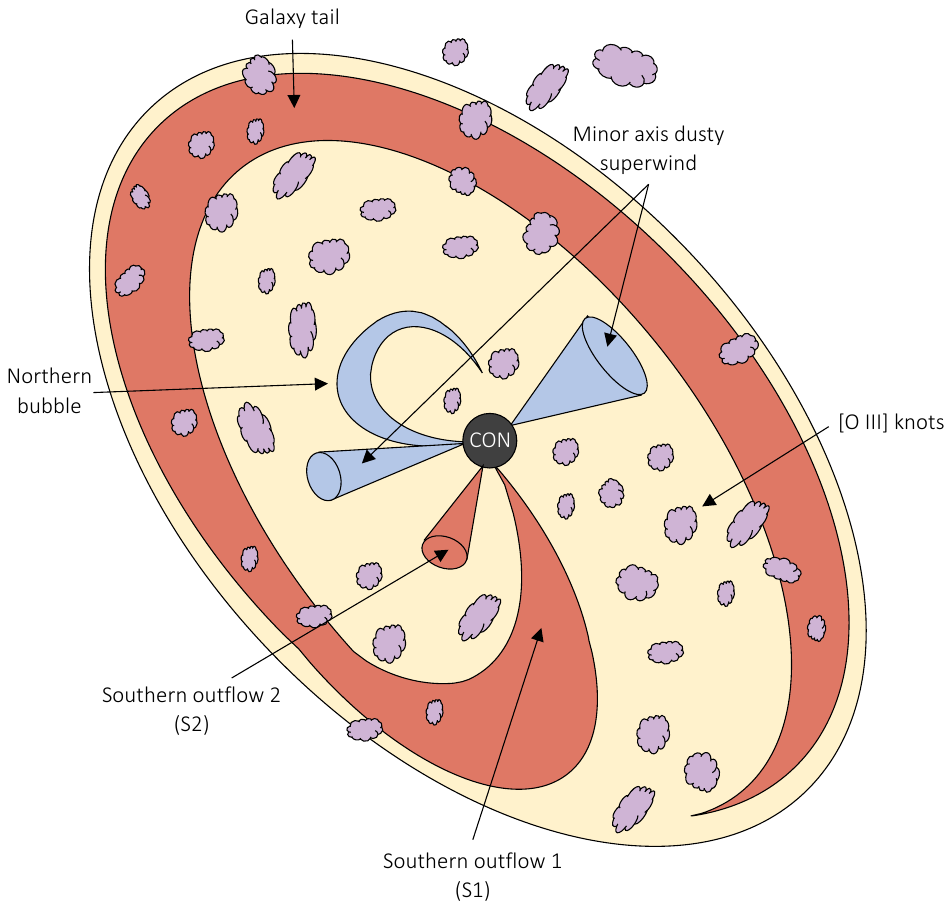}    
        \caption{Schematic illustrating the key features of NGC\,4418 identified in this work.}
    \label{fig:schematic}
    \end{figure}

   In this work we identify five new features in NGC\,4418, which we summarise here as follows: (1) An outflow along the minor axis of the galaxy, extending in both directions from the galaxy nucleus. Throughout this work, we refer to this feature as the {minor axis dusty superwind}. (2) A bubble-like structure to the north, which we refer to as the {northern bubble}. (3) Two outflow features to the south, which we denote as S1 and S2. (4) Knots tracing regions of ionised gas in the galaxy. These are predominantly traced by [O\,{\sc iii}], but a subset are also traced by H$\alpha$. (5) A tail of redshifted material surrounding the galaxy, which appears to originate from the S1 outflow. We name this the {H$\alpha$ tail}. Fig.~\ref{fig:schematic} illustrates the location of these features in NGC\,4418 for reference. Throughout this work, we explore these features in more detail, outlining their key properties and using them to infer information about the CON in NGC\,4418 and its power source.

    %-----------------------------------------------------------------------
    \subsection{Spectral features}
    \label{sec:spectral_features}
    %-----------------------------------------------------------------------
    
        %...................................................................
        \subsubsection{Emission lines}
        \label{sec:emission_lines}
        %...................................................................

        %\begin{figure*}
      %         \centering 
        %    \includegraphics[trim= 170 150 80 150,clip,width=\textwidth]{figs/linemaps.pdf}    
       %     \caption{Extinction-corrected line maps for the emission features of NGC\,4418. (a) H$\beta$ ($\lambda$4861), (b) [O\,{\sc iii}] ($\lambda$5007), (c) [O I]($\lambda$6300), (d) [N\,{\sc ii}] ($\lambda$6548), (e) H$\alpha$ ($\lambda$6563), (f) [N\,{\sc ii}] ($\lambda$6583), (g) [S II] ($\lambda$6716) and (h) [S II] ($\lambda$6731). North is up, east is to the left.}
       % \label{fig:linemaps}
       % \end{figure*}

        \begin{figure*}
        \centering 
            \includegraphics[trim= 80 150 160 140,clip,width=\textwidth]{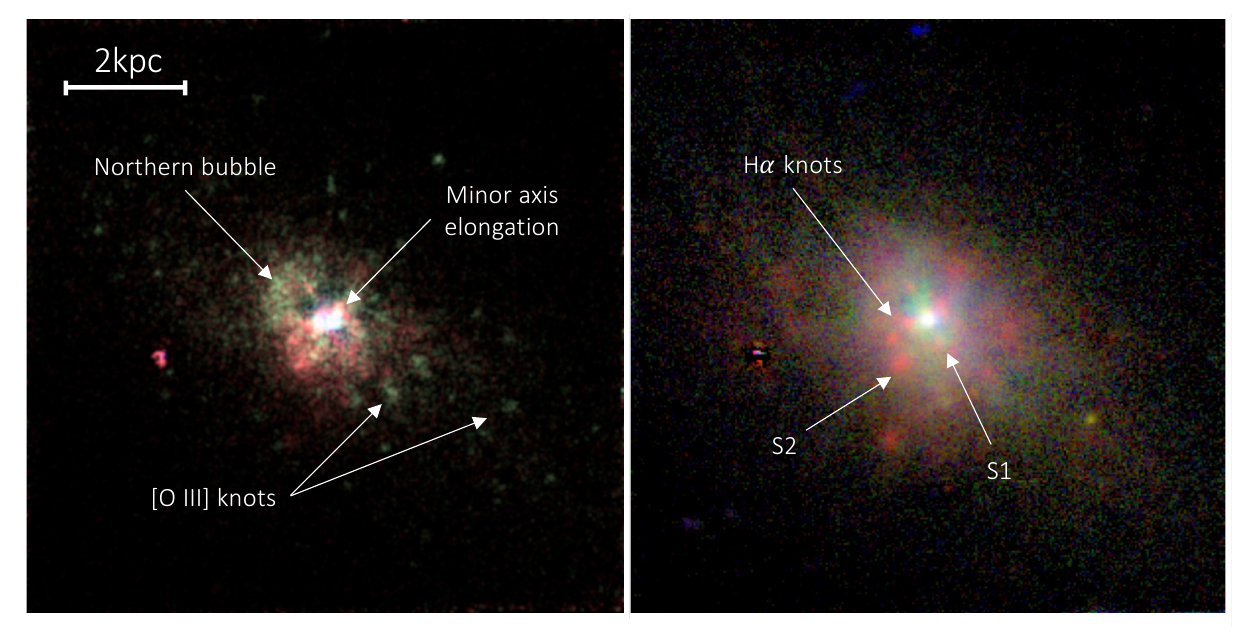}    
            \caption{RGB colour images for \textit{left:} H$\beta$ (red), [O\,{\sc iii}] (green), and [O I] (blue) and  \textit{right:} H$\alpha$ (red), the [N\,{\sc ii}] doublet (green), and the [S II] doublet (blue) with key features labelled. North is up, and east is to the left.}
        \label{fig:linemaps}
        \end{figure*}
        
        In this section we focus on six key diagnostic emission lines tracing the ionised gas in NGC\,4418: H$\beta$ ($\lambda$4861), [O\,{\sc iii}] ($\lambda$5007), [O I]($\lambda$6300), [N\,{\sc ii}] ($\lambda\lambda$6548,6583), H$\alpha$ ($\lambda$6563), and [S II] ($\lambda\lambda$6716,6731). The effect of line blending in the MUSE data cube is negligible, allowing for the effective separation of these different emission lines via channel selection alone. Fig.~\ref{fig:linemaps} presents the Lupton red, green, blue (RGB) colour images for H$\beta$, [O\,{\sc iii}] and [O I], and [N\,{\sc ii}], H$\alpha,$ and [S II] \citep{lupton04}. The component line maps are created by summing the flux across the range of channels associated with each line to create flux-integrated images. These images are then weighted in order optimise the visual output of the resulting colour image Fig.~\ref{fig:linemaps}. This method was chosen to minimise the flux disparity between structures traced by intrinsically brighter (e.g. H$\alpha$) and fainter (e.g. [S II]) lines and thus to best illustrate the key features identified in NGC\,4418. The line emission generally shows an extension along the direction of the minor axis (south-east to north-west), extending in both directions from the galaxy nucleus. In addition to this minor axis structure, we highlight several key features of NGC\,4418 traced by individual emission lines as follows;

        \textbf{H$\beta$ ($\lambda$4861):} We note the presence of two emission features towards the south: one to the south-west, aligning with the major axis of the galaxy, and the other to the south-east. In the interests of clarity, we denote these two southern features as S1 and S2, respectively. In addition to these features, we identify a void to the north-east, surrounded by a shell- or bubble-like emission structure. This structure is reminiscent of that observed in the so-called {teacup galaxy}, J1430+1339 (Treister et al. \textit{in prep.}), which has previously been proposed as a fading AGN candidate \citep{gagne14}. Hereafter, we refer to this feature as the {northern bubble}.
    
        \textbf{[O\,{\sc iii}] ($\lambda$5007):} We identify an emission structure towards the galaxy centre, tracing both the southern features (S1 and S2) and the northern bubble. In addition, the [O\,{\sc iii}] emission reveals bright clumps, or {knots}, throughout the galaxy, which have never before been observed in NGC\,4418. Several of these knots appear extended, yet the majority appear compact and trace a seemingly random and fragmented structure throughout the galaxy. The nature of these knots is explored in detail in section~\ref{sec:the_OIII_knots}.
        
        \textbf{[O I]($\lambda$6300):} The [O I] emission appears visually less extended than that of both H$\beta$ and [O\,{\sc iii}]. [O I] shows a more rapid radial light depletion than either H$\beta$ or [O\,{\sc iii}], indicating a smaller half-light radius. While we identify prominent [O I] emission tracing the S1 feature, we find no evidence that [O I] traces either the S2 feature or the northern bubble.

         \textbf{H$\alpha$ ($\lambda$6563):} We discover several knots traced by the H$\alpha$ emission. Although similar to those identified in the [O\,{\sc iii}] emission, they are fewer in number and generally appear more spatially extended. Towards the centre of the galaxy, a handful of these knots trace the S1 and S2 features, as well as the westward edge of the northern bubble. At larger radii, the knots appear coincident with a small subset of the knots traced by [O\,{\sc iii}].
         
         \textbf{[N\,{\sc ii}] ($\lambda\lambda$6548,6583):} The [N\,{\sc ii}] doublet also traces a bright knot to the south-west of the nucleus, aligning with the S1 feature. We note the presence of an additional, isolated south-western knot further out in the galaxy disk, along with a diffuse knot to the south-east, coinciding with the locations of two of the knots traced by both H$\alpha$ and [O\,{\sc iii}]. With the exception of these two knots, we find no evidence that [N\,{\sc ii}] is enhanced in the knot structures traced by [O\,{\sc iii}]. While we identify faint [N\,{\sc ii}] emission tracing the northern bubble, this feature is much less pronounced than in H$\beta$, [O\,{\sc iii}] or H$\alpha$.
         
         \textbf{[S II] ($\lambda\lambda$6716,6731):}  The [S II] emission doublet traces both the S1 and S2 features, but not the northern bubble. With the exception of two bright knots to the far north, we find no evidence that [S II] is enhanced in any of the knots traced by [O\,{\sc iii}].

         While the features identified above may be tracing outflowing or inflowing material, it is also possible that the emission is instead mapping dust obscuration and/or star forming regions within the galaxy. In order to distinguish between these scenarios and develop a physical interpretation of the line emission observed in NGC\,4418, we therefore require additional kinematic information. In section~\ref{sec:ionised_gas_velocities} we analyse the kinematics of both the stellar disk and the ionised gas traced by the line emission in Fig.~\ref{fig:linemaps}. For now however, we note only the presence of several prominent structures in the line emission.

        %...................................................................
        \subsubsection{The NaD$\lambda\lambda$5890,5896 \r{A} absorption doublet}
        \label{sec:the_nad_absorption_doublet}
        %...................................................................

\begin{figure}
        \centering 
                \includegraphics[trim= 5 5 35 35, clip, width=.5\textwidth]{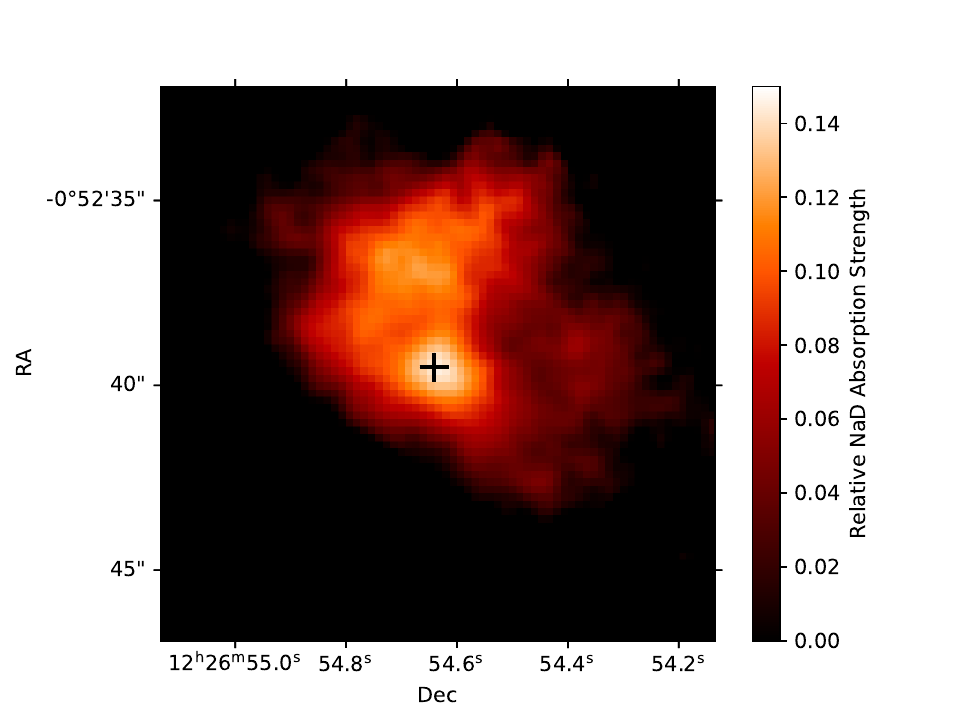}
            \caption{Integrated line map of the NaD$\lambda\lambda$ 5890,5896 \r{A} absorption doublet normalised to the galaxy continuum level.  Colours have been inverted such that stronger absorption corresponds to brighter regions on the map. Black cross denotes the location of the CON.}
        \label{fig:NaD_scaled}
        \end{figure}

        The NaD$\lambda\lambda$5890,5896 $\AA$ absorption doublet traces two distinct components of the interstellar medium (ISM): cold molecular gas and neutral atomic gas. Although the NaD absorption observed in NGC\,4418 is a sum of both these ISM components, the stellar templates used to fit the continuum (see section~\ref{sec:continuum_subtraction}), include only the NaD absorption arising in stellar photospheres. Here, we separate the two ISM components traced by NaD by masking the absorption doublet in the spectra when fitting the continuum. In this way, the stellar component of the NaD absorption included in the stellar templates is constrained only by the continuum strength. Subtracting the best-fit continuum model therefore removes only the absorption due to stellar photospheres and isolates the NaD absorption from neutral atomic gas in the resulting spectra. This remaining NaD absorption is normalised to the galaxy continuum to remove the degeneracy between the strength of the continuum and the absorption depth (i.e. a brighter continuum has a higher absorption potential). Fig.~\ref{fig:NaD_scaled} shows the continuum-normalised NaD absorption, tracing the neutral gas in NGC\,4418. The peak of the absorption coincides with the position of the CON in NGC\,4418 and extends towards both the north and the west, forming a `U' shape.

        %...................................................................
        \subsubsection{H$\beta$ absorption}
        \label{sec:hbeta_absorption}
        %...................................................................

\begin{figure}
        \centering
            \includegraphics[trim= 5 5 20 30,clip,width=.5\textwidth]{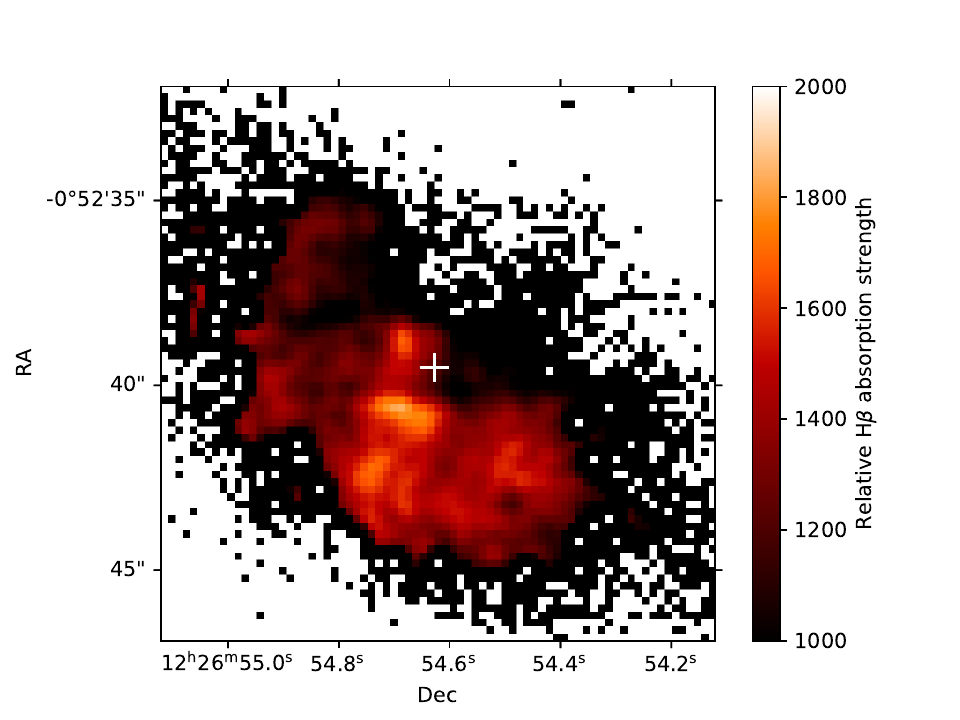}
            \caption{Map of H$\beta$ absorption strength, normalised to the continuum level in each individual spaxel. A 3$\sigma$ S/N cut has been applied and the image has been smoothed by a 1$\sigma$ Gaussian kernel to improve visual output. Colours have been inverted such that stronger absorption corresponds to brighter regions on the map. White cross denotes the location of the CON.}
        \label{fig:Hb_abs_norm_map}
        \end{figure}

        The equivalent width of the H$\beta$ absorption is highly sensitive to the initial mass function (IMF) of the galaxy and thus is often used as an indicator of the underlying stellar population \citep[e.g.][]{copetti86}, primarily tracing the presence of A-type stars in the galaxy. To explore the stellar properties of NGC\,4418, we therefore map the H$\beta$ absorption across the galaxy. To this end, we perform a spaxel-by-spaxel non-linear least squares fitting to the original data cube, prior to any continuum modelling and subtraction. The region of the spectrum covering the H$\beta$ absorption feature (4750 - 4950 \r{A}) is isolated and each spaxel is fit with a two-component Gaussian model, consisting of both a narrow emission and a broad absorption component. The continuum strength is estimated from a nearby, featureless region of the spectrum and is also included in the fitting. A signal-to-noise ratio cut of three is applied prior to this fitting such that we consider only spaxels for which the H$\beta$ absorption strength exceeds three times the standard deviation of the spectral noise. The total H$\beta$ absorption in each spaxel is calculated by integrating over the broad absorption component of the best-fit model. As in section~\ref{sec:the_nad_absorption_doublet}, the resulting H$\beta$ absorption is then normalised to the strength of the continuum and is plotted in Fig.~\ref{fig:Hb_abs_norm_map}, where we reveal several notable features in the H$\beta$ absorption of NGC\,4418. We identify a strong absorption feature to the south-east of the galaxy nucleus, which appears elongated along the east-west axis. We additionally find an absorption feature to the north-east, which we suggest is associated with the void of the bubble structure seen on the H$\beta$ emission (Fig.~\ref{fig:linemaps}). To the south, we also find an enhancement in the absorption strength of H$\beta$, potentially associated with the S1 and S2 features.
     
    %-----------------------------------------------------------------------
    \subsection{Dust extinction}
    \label{sec:dust_extinction}
    %-----------------------------------------------------------------------

    Having mapped the ionised (emission) and neutral (absorption) components of the gas in NGC\,4418, we now map the dust extinction in the galaxy. The total extinction, $A_v$, is calculated from the Balmer decrement of the continuum-subtracted spectra (Eq.~\ref{eqn:balm_dec}),

        \begin{equation}
        \label{eqn:balm_dec}
            \frac{F(H\alpha)}{F(H\beta)} = \frac{F_0(H\alpha)}{F_0(H\beta)} 10^{\frac{-0.4A_v(k(H\alpha) - k(H\beta))}{R_V}},
        \end{equation}

    \noindent
    where $F$ is the peak flux of the given emission line, R$_V$ is the effective total obscuration in the V-band and $k$ is the chosen dust extinction law. During the line fitting, the width of the emission features, $\sigma$, were fixed to be the same, meaning the peak flux ratio is equivalent to the ratio of the integrated fluxes of the lines. We assume a \cite{calzetti00} attenuation law, which gives R$_V$=4.05 and $k(H\alpha)-k(H\beta)$=-0.58. This law has been empirically derived from nearby starburst galaxies and thus provides a suitable approximation for NGC\,4418. Based on the work of \cite{osterbrock06}, we assume an intrinsic flux ratio:
    
        \begin{equation}
        \label{eqn:flux_rat}
            \frac{F_0(H\alpha)}{F_0(H\beta)} = 2.86.
        \end{equation}

\begin{figure}
    \centering
            \includegraphics[trim= 0 0 40 30,clip,width=.5\textwidth]{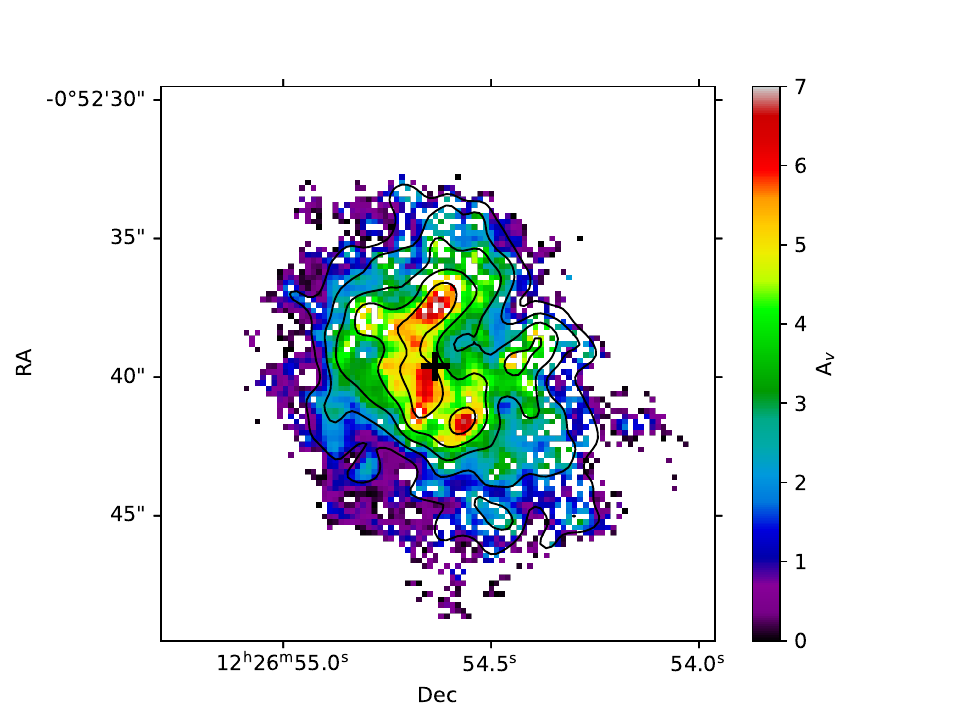}
            \caption{Map of $A_v$ values, derived from the Balmer decrement. The map has been smoothed with a 1$\sigma$ Gaussian kernel to improve the visual output and a signal-to-noise ratio cut of three has been applied.}
    \label{fig:dust_balmer}
    \end{figure}

    The value of $A_v$ for each spaxel is plotted in Fig.~\ref{fig:dust_balmer}, where a signal-to-noise ratio cut of three has been imposed to remove any spurious measurements. The resulting extinction map (Fig.~\ref{fig:dust_balmer}) shows a similar 'U' shape to that observed in the NaD neutral gas (Fig.~\ref{fig:NaD_scaled}), with an extended structure to the north. Furthermore, we identify two prominent dust structures to the south, aligning with both the S1 and S2 emission line features. We note the lack of dust along the direction of the minor axis of the galaxy, implying the dust in this region may have been cleared out by, for example, an outflow.

    %-----------------------------------------------------------------------
    \subsection{Galaxy dynamics}
    \label{sec:galaxy_dynamics}
    %-----------------------------------------------------------------------

    Although the line emission and absorption features in NGC\,4418 provide valuable information on the morphological structure, stellar populations and dust composition within the galaxy, many of the structures we observe cannot be physically interpreted without dynamical information. To this end, we exploit the IFU capabilities of MUSE to present the kinematic properties of NGC\,4418. 
    
        %...................................................................
        \subsubsection{Stellar kinematics}
        \label{sec:stellar_kinematics}
        %...................................................................

\begin{figure}
        \centering
            \includegraphics[trim= 125 290 150 320,clip,width=.5\textwidth]{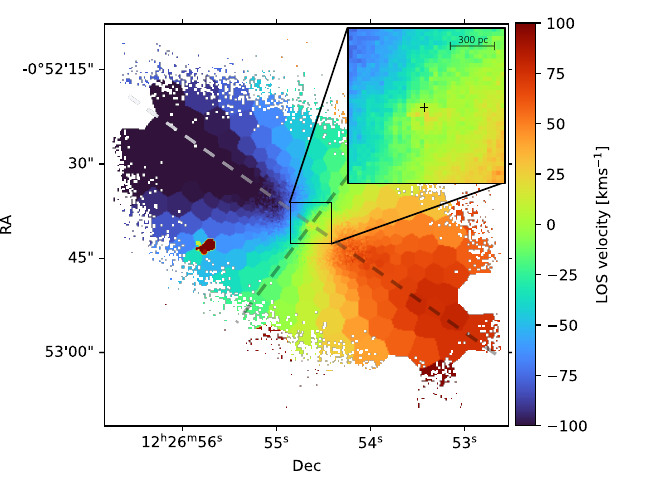}
            \caption{Map of the stellar velocities derived from the pPXF fitting. The black square denotes the extent of the 7$\times$7 arcsec$^2$ zoom-in region shown in the insert. The position of the CON is marked with a black cross and dashed lines mark the positions of the major and minor axes. Pixels have been binned to achieve a signal-to-noise ratio of 20. An initial signal-to-noise ratio cut of 2 has been applied, with pixels below this threshold omitted from the binning.}
        \label{fig:stel_vel_binned}
        \end{figure}

        The rotation of the stellar disk of NGC\,4418 is mapped from the wavelength shift of the best-fit stellar templates during the continuum modelling (section~\ref{sec:continuum_subtraction}). During this modelling process, all prominent emission lines in the spectra are masked, meaning the fitted stellar templates are constrained only by the spectral absorption. While the stellar velocities can therefore be measured for each spaxel in the cube individually, we instead opt to bin the data into regions with a fixed signal-to-noise. To achieve this, we implement a Voronoi adaptive binning system \citep{cappellari03}, using the 5000$\AA$ flux as a reference for the binning. The continuum modelling was re-run on the binned data cube, with a target signal-to-noise ratio of 20 in each bin. Fig.~\ref{fig:stel_vel_binned} maps the resulting stellar velocities derived in each bin. We discover an asymmetric disk rotation in NGC\,4418, with the north-east side approaching with velocities, V$\sim$-115 km\,s$^{-1}$ and the south-west side receding with V$\sim$80 km\,s$^{-1}$. Such asymmetries can be caused by a major tidal disruption event, such as a major merger. It is therefore possible that NGC\,4418 underwent such an event in the past, potentially with the nearby galaxy VV655 \citep[e.g.][]{boettcher20}, although we note that we see no conclusive evidence for this.
     
\begin{figure}
        \centering
            \includegraphics[trim= 0 10 40 55, clip, width=.45\textwidth]{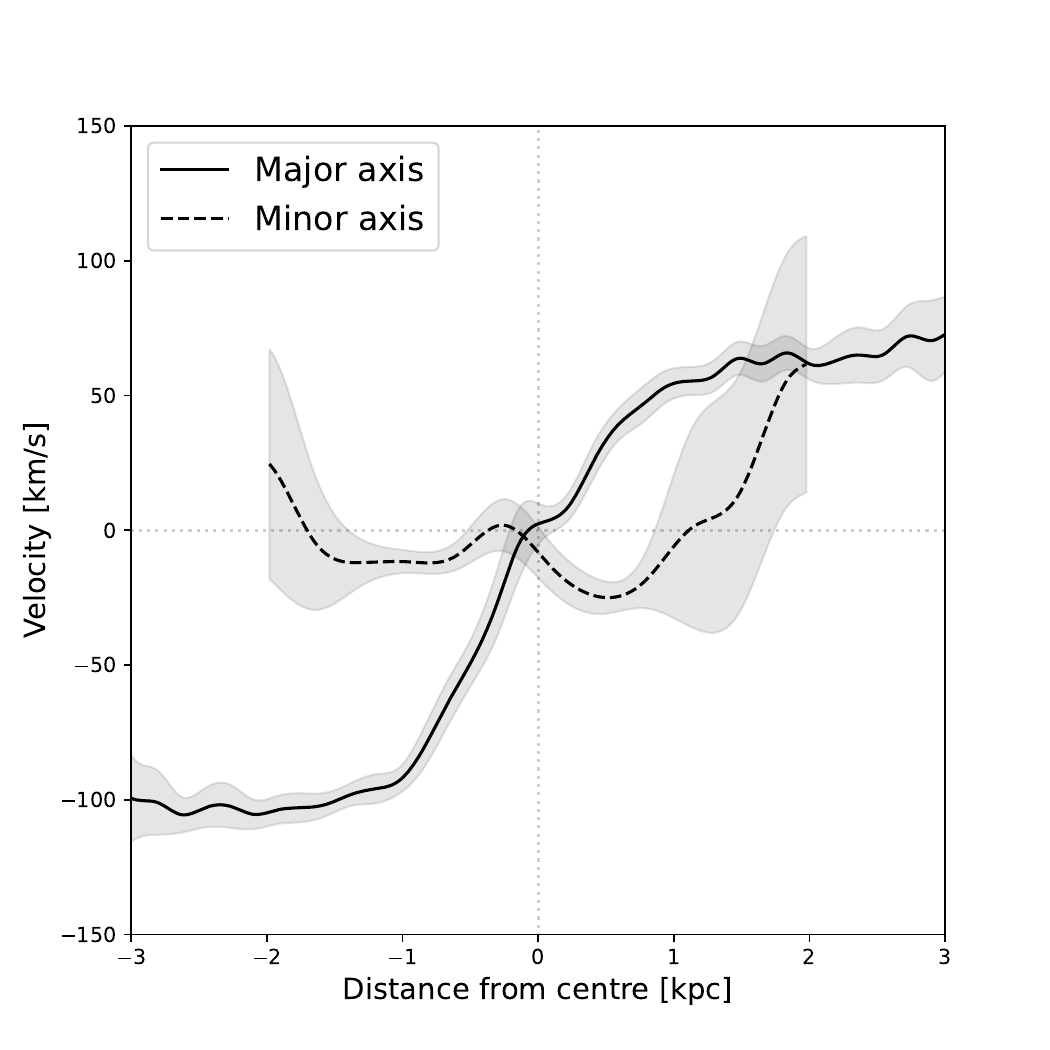}           
            \caption{Galaxy rotation curves along the kinematic major (solid line)  and minor (dashed line) axes of NGC\,4418. In each case, velocities are averaged over a simulated slit with a width of $\pm$3 pixels ($\pm$0.6\, arcsec). Shaded regions denote a standard deviation from the average velocity.}
        \label{fig:rot_curve}
        \end{figure}

        Within the central kiloparsec we identify a structure that does not appear to be rotating with the stellar disk. Although difficult to resolve due to the spatial resolution of MUSE (0.4 arcsec), the velocities in this central region appear counter-aligned with the major axis rotation of the galaxy. Given the region's proximity to the galaxy centre, the peculiar velocities can be seen in both the major and minor axis rotation curves of NGC\,4418 (Fig.~\ref{fig:rot_curve}). Along the major axis (Fig.~\ref{fig:rot_curve}), we identify a small `bump' in the velocity structure. Likewise, the minor axis rotation curve (Fig.~\ref{fig:rot_curve}) also shows a `bump', offset by $\sim$0.25 kpc from the galaxy centre. 
        
        We suggest that the anomalous velocity structure we observe may indicate the presence of a counter-rotating nuclear disk. This conclusion is consistent with the results of \cite{ohyama19}, who also find evidence for a counter-rotating structure in the centre of NGC\,4418 based on single-slit spectroscopy. We note however that \cite{ohyama19} do not explicitly distinguish between a counter-rotating gas disk and other non-rotational motions of the central gas. Despite the lack of a large-scale bar in NGC\,4418 we suggest that the peculiar velocity structure may alternatively mark the presence of a nuclear bar. One test to help distinguish between these scenarios is to look for any (anti-)correlation in the rotation velocities of the stellar disk and the third Gauss-Hermite coefficient, h3. The h3 parameter is output from the pPXF fitting described in section~\ref{sec:continuum_subtraction}. An anti-correlation between these parameters would indicate the likely existence of a counter-rotation as opposed to a nuclear bar. Indeed, we observe an anti-correlation between the h3 parameter and the rotation of the galaxy disk (Fig.~\ref{fig:veldisp_h3}), with the positive (red skewed) h3 coefficients to the north-east (blue side) of the stellar disk. While we emphasise that the spatial resolution of MUSE is insufficient to map the detailed kinematics of the central kiloparsec, the broad anti-correlation between the h3 coefficient and the rotation of the galaxy disk supports our conclusion of a counter-rotation. 

       We additionally identify a clear enhancement in the stellar velocity dispersion, $\sigma$, towards the centre of NGC\,4418 (Fig.~\ref{fig:veldisp_h3}). This enhanced dispersion forms a small cross-like structure extending both along the minor axis of the galaxy, coincident with the location of the ionised outflow, and along the major axis tracing the galaxy disk rotation. Due to the orientation of this feature in, we suggest it is likely that the enhanced velocity dispersion is associated with the outflow and galaxy disk. It is also possible however, that we are instead observing two misaligned nuclear disks; a reasonable scenario given the merger history of NGC\,4418. While the limited resolution of the MUSE imaging prevents us from robustly distinguishing between these scenarios, observations with ALMA at higher spatial resolution confirm that the nuclear gas disk in NGC\,4418 is counter-rotating with respect to the galaxy disk \citep{sakamoto21}. Although \cite{sakamoto21} attribute this feature to be a disturbance resulting from an outflow, nuclear rotation, or turbulence, \cite{varenius17} also suggest gas inflow from a companion galaxy as a potential explanation. Regardless of the exact nature of the central velocity structure observed in NGC\,4418, it is a clear indicator of a major ongoing or recent disturbance to the nuclear region, perhaps associated with a recent merger or interaction event.
        
        %...................................................................
        \subsubsection{Ionised gas velocities}
        \label{sec:ionised_gas_velocities}
        %...................................................................

\begin{figure*}
            \centering 
                \includegraphics[trim= 10 190 60 150,clip,width=\textwidth]{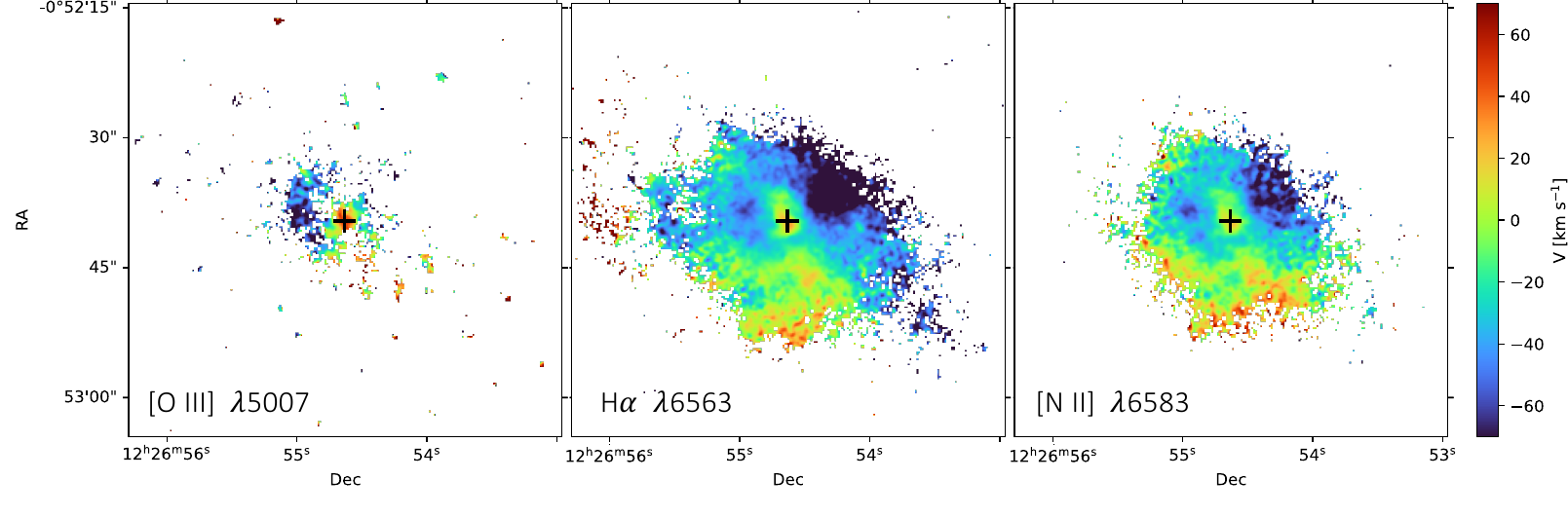}             
                \caption{Moment-1 velocity maps of the ionised gas traced by [O\,{\sc iii}] (left), H$\alpha$ (centre), and [N\,{\sc ii}]$\lambda$6583 (right). In each case, a signal-to-noise ratio cut of three has been applied. Maps have been smoothed by a $\sigma$ = 1 spaxel Gaussian kernel to improve visual output.}
         \label{fig:ionised}
         \end{figure*}

        Fig.~\ref{fig:ionised} shows the moment-1 velocity maps for [O\,{\sc iii}] ($\lambda$5007), H$\alpha$ and [N\,{\sc ii}] ($\lambda$6583), tracing the motions of the ionised gas in NGC\,4418. The moment maps were created using the Scipy package \textsc{scipy.stats.moment}, and have undergone a signal-to-noise ratio cut of three to remove any spurious velocity features. All three tracers show clear non-circular motions, independent from the rotation of the stellar disk. In each case, we identify a blueshifted structure to the east of the galaxy nucleus with a peak velocity, $V\sim$75 km\,s$^{-1}$. In the moment-1 maps of H$\alpha$ and [N\,{\sc ii}] (Fig.~\ref{fig:ionised}), this eastern blueshifted structure comprises multiple distinct regions, appearing as three separate structures in the H$\alpha$ velocity map and two in the [N\,{\sc ii}] map. In both cases, the northern-most structure is the most prominent. This northern-most blueshifted structure appears spatially aligned with the void of the northern bubble seen in the H$\beta$ line emission (section~\ref{sec:emission_lines}). The moment-1 maps of H$\alpha$ and [N\,{\sc ii}] also reveal a blueshifted feature to the west of the nucleus with velocities up to $V\sim$140 km\,s$^{-1}$. This feature is much larger than that to the east, extending $>$1 kpc from the galaxy centre, and appears aligned with the minor axis elongation traced by the emission lines (Fig.~\ref{fig:linemaps}). 

        In addition to these blueshifted structures, we detect a slower ($V<$50 km\,s$^{-1}$) redshifted feature to the south. This feature is present in all three of the velocity maps in Fig.~\ref{fig:ionised}, albeit faint in [O\,{\sc iii}]. This redshifted structure appears aligned with the S1 feature we identify to the south. In the H$\alpha$ and [N\,{\sc ii}] velocity maps, there is some indication that this structure also traces the S2 feature, but this feature is comparatively faint in terms of its velocity shift. We therefore conclude that the S2 feature is likely tracing slower material than S1. 

        %...................................................................
        \subsubsection{Velocity dispersion of the ionised gas}
        \label{sec:velocity_dispersion}
        %...................................................................

\begin{figure*}
            \centering 
                \includegraphics[trim= 10 190 75 150,clip,width=\textwidth]{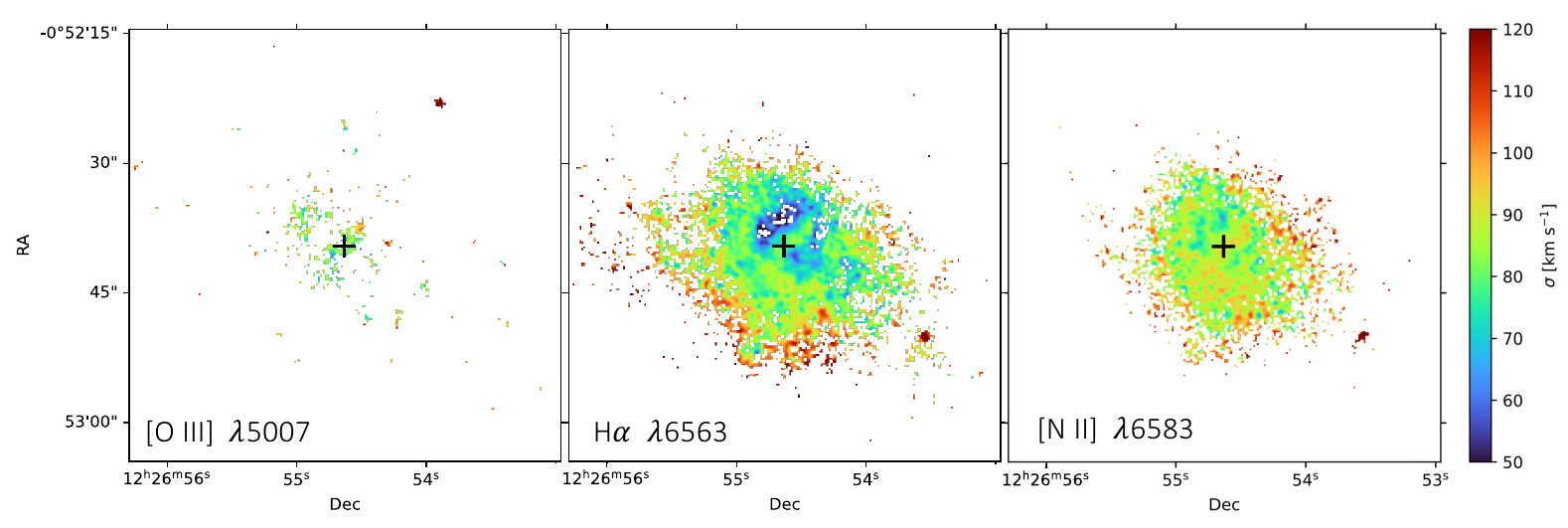}             
                \caption{Moment-2 velocity dispersion maps of the ionised gas traced by [O\,{\sc iii}] (left), H$\alpha$ (centre), and [N\,{\sc ii}]$\lambda$6583 (right). A signal-to-noise ratio cut of three has been applied. Maps have been smoothed by a $\sigma$ = 1 spaxel Gaussian kernel to improve visual output. North is up, east is to the left.}
         \label{fig:vel_disp}
         \end{figure*}

        Fig.~\ref{fig:vel_disp} shows the moment-2 velocity dispersion maps for [O\,{\sc iii}] ($\lambda$5007), H$\alpha$ and [N\,{\sc ii}] ($\lambda$6583), tracing the random motions of the ionised gas along the line of sight and in turn probing the motions of the stars in the gravitational potential of the galaxy. The  H$\alpha$ and [N\,{\sc ii}] maps (Figs.~\ref{fig:vel_disp}) show an inverted `3' shaped structure, angled towards the north-west, over which the velocity dispersion is enhanced. The central part of this structure appears aligned with the extension along the minor axis we observe in the line emission (Fig.~\ref{fig:linemaps}) and with the strongly blueshifted structure seen in the moment-1 velocity maps (Fig.~\ref{fig:ionised}). If this blueshifted structure is an outflow, we therefore expect it to contain a significant amount of mass. The regions at either side of this minor axis structure have a lower velocity dispersion, implying that either the mass in these regions has been cleared out or the mass is in some way collimated along the minor axis of NGC\,4418. We also identify an enhancement in the velocity dispersion towards the south, in apparent alignment with the redshifted S1 and S2 features. 

    %-----------------------------------------------------------------------
    \subsection{Inflows and outflows}
    \label{sec:inflows_and_outflows}
    %-----------------------------------------------------------------------

\begin{figure}
            \centering
                \includegraphics[trim= 120 30 170 40,clip,width=.45\textwidth]{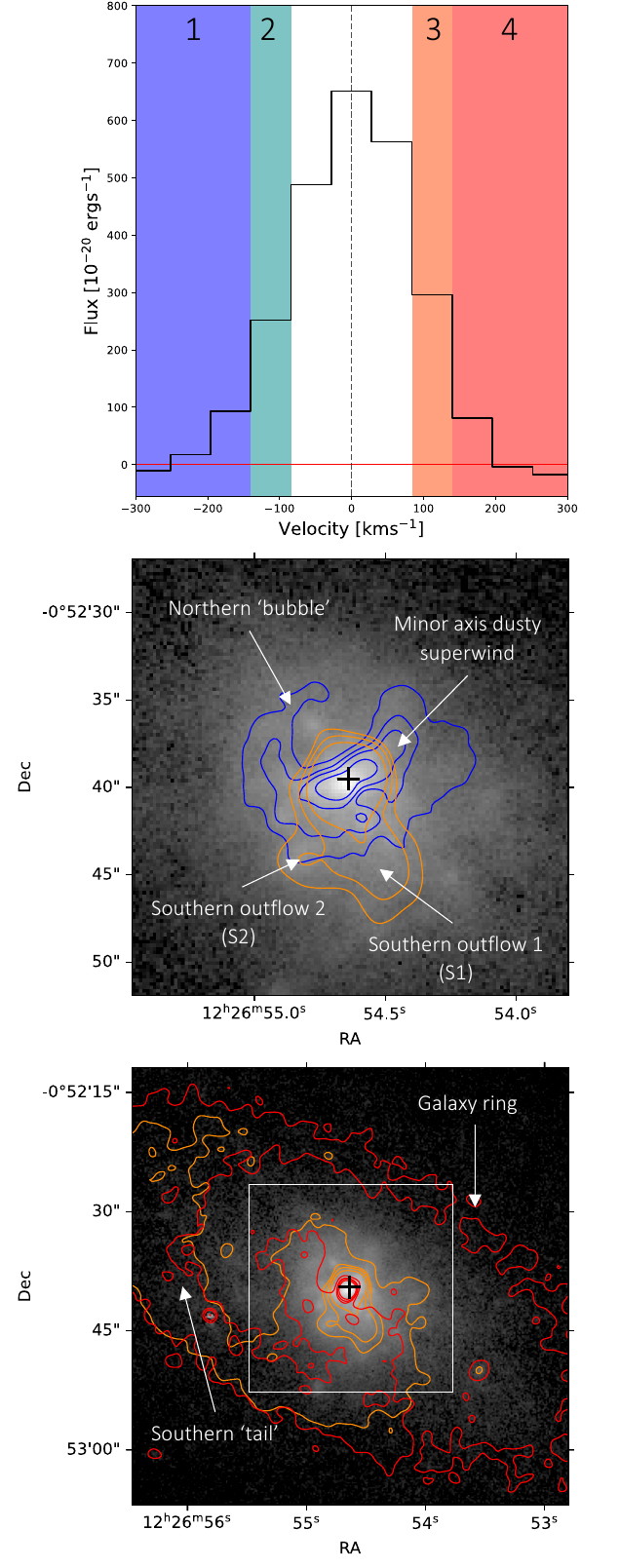}
                \caption{Emission features in the H$\alpha$ line wings. \textit{Upper:} Spectrum of H$\alpha$, extracted from a 2.5 arcsec (diameter) aperture centred on the CON position. Blue and red shaded regions are overlaid and labelled from 1-4. Regions 1 and 4 contain the highest velocity (140$<V>$305 km\,s$^{-1}$) blue- and redshifted gas, respectively. Regions 2 and 3 map include the slower (85$<V>$140 km\,s$^{-1}$) gas.
                \textit{Middle:} Emission of the fast blueshifted (region 1; blue contours) and slower redshifted (region 3; orange contours) H$\alpha$ gas, overlaid onto the integrated H$\alpha$ line map (log-scaled). 
                \textit{Lower:} Emission of the fast (region 4; orange contours) and slower (region 3; red contours) redshifted H$\alpha$ gas. White box shows the region shown in the middle panel.}
            \label{fig:wings}
        \end{figure}

        To explore the outflow and inflow structures in NGC\,4418 in more detail, we separate the H$\alpha$ line emission into different velocity regions. During the continuum modelling, all spectra in the MUSE cube were re-binned such that each channel has a fixed velocity width of $\Delta$V = 55.9 km\,s$^{-1}$. Based on these velocities, we separate the H$\alpha$ emission into four velocity regions, tracing the gas at different velocities (Fig.~\ref{fig:wings}). Regions 1 and 4 trace the highest velocity gas (140 $< v <$ 300 km\,s$^{-1}$) on the blue and red wings, respectively, while regions 2 and 3 trace the slightly slower gas (85$< v <$140 km\,s$^{-1}$). By isolating the emission within each of these different velocity regions, we are able to spatially map the high- and intermediate- velocity gas in NGC\,4418 for the first time, to reveal several new structures in the galaxy. Combined with the kinematic information presented in section~\ref{sec:galaxy_dynamics}, we identify four key features: 1. the minor-axis dusty superwind, 2. the northern bubble, 3. the southern outflow and 4. the H$\alpha$ 'tail'. 

        %...................................................................
        \subsubsection{The dusty superwind}
        \label{sec:the_dusty_superwind}
        %...................................................................

        The blueshifted high velocity gas (region 1 in Fig.~\ref{fig:wings}; $V<$-140 km\,s$^{-1}$) traces a notable asymmetry along the minor axis of NGC\,4418. The position of this asymmetry is consistent with the location of both the strongly blueshifted regions traced by the H$\alpha$ moment-1 velocity map (Fig.~\ref{fig:ionised}) and the region of enhanced velocity dispersion (Fig.~\ref{fig:vel_disp}). We attribute this feature to a large-scale ($>$1 kpc) outflow along the direction of the minor axis. The existence of such a large-scale outflow in NGC\,4418 was first proposed by \cite{sakamoto13}, who observed a 'U'-shaped region of heavy extinction along the minor axis of the galaxy in their optical colour map. This 'U'-shaped region is very similar to the structure we observe in the NaD absorption and the reddening map with MUSE (Figs. ~\ref{fig:NaD_scaled} and ~\ref{fig:dust_balmer}, respectively). \cite{sakamoto13} proposed that this dust structure was likely a result of an outflow cone, appearing aligned with the minor axis of the galactic disk on the plane of the sky. Indeed, follow up observations by \cite{ohyama19} also found evidence for this structure based on slit spectroscopy from the Subaru telescope. They concluded that NGC\,4418 hosts an outflow, or {dusty superwind}, launched from the nucleus and extending in the north-west direction with a velocity of $\sim$-190 km\,s$^{-1}$. We suggest that the minor-axis structure observed here with MUSE is likely linked to this outflow cone, or dusty superwind, postulated by both \cite{sakamoto13} and \cite{ohyama19}. The position of the outflow structure seen in the high velocity channels of H$\alpha$ aligns with the inner section of the 'U' shape seen in the dust. Our results therefore support a scenario in which much of the dust and neutral gas has been expelled from this region of the galaxy via this superwind, or in which this is part of an outflowing cone structure.
        
        In addition to the previously postulated north-west component of this outflow however, we also discover a blue component to the SE. We therefore suggest that the outflow observed in NGC\,4418 does not just extend to the north-west, but rather on either side of the nucleus. The high-velocity blueshifted emission tracing this outflow feature is shown in the blue contours of Fig.~\ref{fig:wings}, where we identify a clear elongation along the projected minor axis of the galaxy. The outflow appears to extend to $\sim$1 kpc in either direction from the nucleus and is blueshifted in both directions. Furthermore, the outflow is found to be decelerating, with the low velocity ($V<$85 km\,s$^{-1}$) gas appearing much more extended than the fastest gas, which is concentrated in the nuclear region and depletes with distance from the galaxy centre.  
        
        %...................................................................
        \subsubsection{The northern bubble}
        \label{sec:the_northern bubble}
        %...................................................................

        In addition to the minor axis superwind, we identify a blueshifted shell- or bubble-like structure to the NE, traced by the highest velocity gas (region 1). This structure, which we refer to as the {northern bubble} appears connected to the SE side of the minor axis superwind and is illustrated in Fig.~\ref{fig:wings}. This northern bubble is spatially aligned with the position of the bubble structure traced by the line emission (Fig.~\ref{fig:linemaps}), particularly H$\beta$. Furthermore, recent \textit{Hubble} Space Telescope (HST) observations of NGC\,4418 (Gallagher et al. \textit{in prep.}) support the presence of such a bubble, showing dust arcs extending northwards of the galaxy centre and tracing a similar structure. Our results show that the fastest gas traces the edge of the bubble, with the slower gas ($V>$-85 km\,s$^{-1}$) tracing clumpy structures in the centre, or void, of the bubble. This suggests that the outflowing bubble may be  accelerating, with the edges of the bubble expanding much faster than the inner, void region. Alternatively, this could be an orientation effect. Given that the outer edge of the bubble appears connected to the minor axis outflow, it is therefore possible they are part of the same structure, oriented in such a way so as to appear as a bubble in the plane of the sky. Some tentative evidence for this scenario is seen in the H$\alpha$ moment-1 map (Fig.~\ref{fig:ionised}), where the northern-most edge of the bubble is shown to be slightly redshifted relative to the rest of the bubble. It is therefore possible that this northern-most edge is instead tracing inflowing material, that is to say, material is being ejected towards the observer via the dusty superwind before falling back onto the plane of the galaxy.  

        %...................................................................
        \subsubsection{The southern outflow}
        \label{sec:the_southern_tail}
        %...................................................................

        The intermediate-velocity redshifted gas (region 3 in Fig.~\ref{fig:wings}; 85 $<V<$ 140 km\,s$^{-1}$) traces a notable feature to the south of the galaxy nucleus in NGC\,4418. This feature appears aligned with the location of both the S1 and S2 emission features identified in the line emission (Fig.~\ref{fig:linemaps}) and moment-1 maps of the ionised gas (Fig.~\ref{fig:ionised}). Given the velocity structure of this feature, we conclude the presence of two redshifted outflows to the south of the galaxy nucleus: a primary outflow, S1, extended on scales $>$1 kpc, and a secondary, more compact outflow, S2. Due to their redshifted velocities, we infer that both outflows are located behind the galaxy disk. However, it is also possible that these structures are instead inflows located in front of this disk.
        
        %...................................................................
        \subsubsection{The H$\alpha$ tail}
        \label{sec:halpha_tail}
        %................................................................... 
        
        On larger scales, the same redshifted intermediate-velocity gas traces a large-scale feature extending south from the nucleus and to the north-east (Fig.~\ref{fig:wings}), which we refer to as the {H$\alpha$ tail}. Moreover, the highest velocity redshifted gas (region 4; $V>$ 140 km\,s$^{-1}$) also traces much of this tail, extending further into a ring shape surrounding the galaxy. While the high velocity ring is much fainter than both the tail and the southern outflow, they appear to be connected. Both the tail and the ring are likely indicative of a past large-scale disturbance to the galaxy, such as a major interaction event. This interpretation would be consistent with the asymmetry in the stellar disk identified in section~\ref{sec:stellar_kinematics}. Alternatively, the tail feature we identify could be part of a spiral structure in the galaxy. Indeed, imaging from the Sloan Digital Sky Survey (SDSS) reveals what appears to be a faint spiral arm originating from the south-west (red) side of the stellar disk and aligning with the tail structure. This could also explain the asymmetry observed in the stellar disk and provides a more elegant explanation for the unusual redshifted tail structure we observe.

%***************************************************************************
\section{Discussion}
\label{sec:discussion}
%***************************************************************************

    %-----------------------------------------------------------------------
    \subsection{Comparing dust extinction diagnostics}
    \label{sec:comparing_dust_extinction_diagnostics}
    %-----------------------------------------------------------------------

\begin{figure}
    \centering
            \includegraphics[trim= 5 5 30 30,clip,width=.5\textwidth]{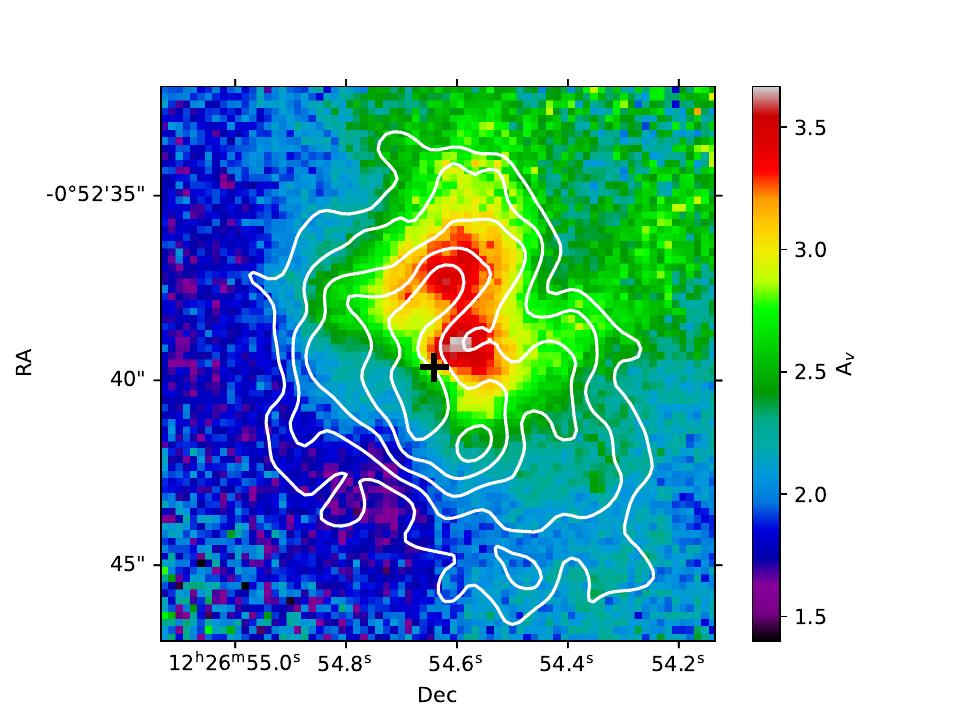}
            \caption{Map of the total dust extinction, $A_{v,\rm{(g-r)}}$, derived from the KiDS g'-r' (AB) colours. Contours mapping $A_{v,\rm{balmer}}$ values derived from the Balmer decrement of the MUSE spectra are overlaid (white).}
    \label{fig:dust_comp}
    \end{figure}

     In section~\ref{sec:dust_extinction} we map the dust structure of NGC\,4418 based on the Balmer decrement. Here, we compare the total dust extinction derived in this work, hereafter $A_{v,\rm{balmer}}$, with that derived from the Kilo-Degree Survey (KiDS) $g'-r'$ colours, $A_{v,\rm{colour}}$, presented in \cite{ohyama19}. We derive $A_{v,\rm{colour}}$ from the $g'-r'$ colours, $E(g'-r')$, presented in \cite{ohyama19}, following the methods outlined in \cite{tian14} (Eq.~\ref{eqn:g_r}):

     \begin{equation}
        \label{eqn:g_r}
            E(g-r) = \left(\left( a_g + \frac{b_g}{R_V} \right) - \left(a_r +  \frac{b_r)}{R_V} \right)\right) \times A_V,
     \end{equation}where the wavelength-dependent coefficients, $a$ and $b$, are calculated for the optical/near-IR (1.1$\leq x \leq$3.3 $\mu$ m$^{-1}$) as follows \citep[Eqns. 3a,b in][]{cardelli89}:

\begin{subequations}
\begin{align}
\begin{split}
a(x) = &1 + 0.17699y - 0.50447y^2 - 0.02427y^3 + 0.72085y^4\\
&+ 0.01979y^5 - 0.77530y^6 + 0.32999y^7
\end{split}\\
\begin{split}
b(x) = &1.41338y + 2.28305y^2 + 1.07233y^3 - 5.38434y^4 \\
 &- 0.62251y^5 + 5.30260y^6 - 2.09002y^7,
\end{split}
\end{align}
\label{straincomponent}
\end{subequations}

   \noindent where $x$ is the wavenumber of the band, $1 / \lambda_{\rm{eff}}$, and $y = x - 1.82$. Fig.~\ref{fig:dust_comp} shows the resulting $A_{v,\rm{colour}}$ map, with contours of the $A_{v,\rm{balmer}}$ overlaid. While both $A_{v,\rm{colour}}$ and $A_{v,\rm{balmer}}$ trace the same characteristic 'U' structure, we note that $A_{v,\rm{balmer}}$ traces an additional prominent dust feature to the south, which is not present in the map of $A_{v,\rm{colour}}$. This feature is spatially coincident with the redshifted southern features, S1 and S2, discussed in this work. We suggest that this discrepancy in the dust distribution within the galaxy is primarily due to the different methods of measuring the dust. Extinction measurements based on ionised emission, such as the $A_{v,\rm{balmer}}$ value we derive, are typically more sensitive to dust behind the disk of the galaxy than $A_{v,\rm{colour}}$, which has been derived from the broad band colours. In principle, the dust map derived from MUSE may therefore be tracing additional features not traced by the KiDS colour map. As such, we suggest that the dusty southern features likely lie behind the plane of the galaxy. Given that we find these features to be kinematically redshifted, we suggest that they are indeed outflowing components as postulated in section~\ref{sec:the_southern_tail}. 

    %-----------------------------------------------------------------------
    \subsection{The [O\,{\sc iii}] knots}
    \label{sec:the_OIII_knots}
    %-----------------------------------------------------------------------

\begin{figure}
        \includegraphics[trim= 30 5 70 35,clip,width=.5\textwidth]{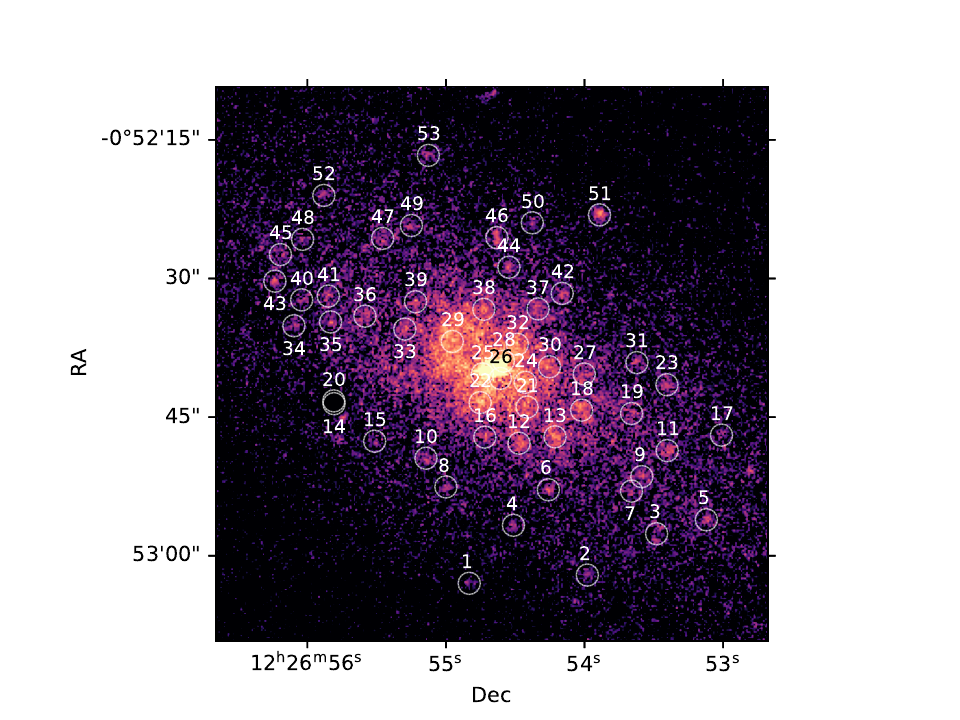}
        \caption{Locations of the [O\,{\sc iii}] knots.}
    \label{fig:knot_loc}
    \end{figure}

    In section~\ref{sec:emission_lines} we identify clumpy structures in the ionised [O\,{\sc iii}] emission, which have never been observed before in NGC\,4418. To further investigate the nature of these ionised regions, or {knots}, we extract the location of each knot by convolving the integrated [O\,{\sc iii}] line map (Fig.~\ref{fig:knot_loc}) with a Gaussian kernel and subtracting it from the original map in a rudimentary high-pass filter. A threshold is applied to this residual image to remove any obvious noise signals and isolate the brightest knots. The threshold value was selected by eye such that all spurious sources in the image are removed and all knots appearing as separate structures under visual inspection remain separate in the threshold image  (i.e. the threshold was sufficiently large to avoid blending multiple nearby knots into a single structure). Contours are fitted to the resulting threshold map and the centre of each contour is chosen to denote the location of the corresponding knot. Through this method, we identify positions for 51 of the brightest knots in NGC\,4418 (K01-K51; Fig.~\ref{fig:knot_loc}), the details of which are given in Table~\ref{tab:a1}. We highlight that many other, fainter structures similar to the bright knots also exist throughout the galaxy, but were too faint to be selected via this method. 

\begin{figure*}
    \centering
        \includegraphics[trim= 90 75 90 70,clip,width=\textwidth]{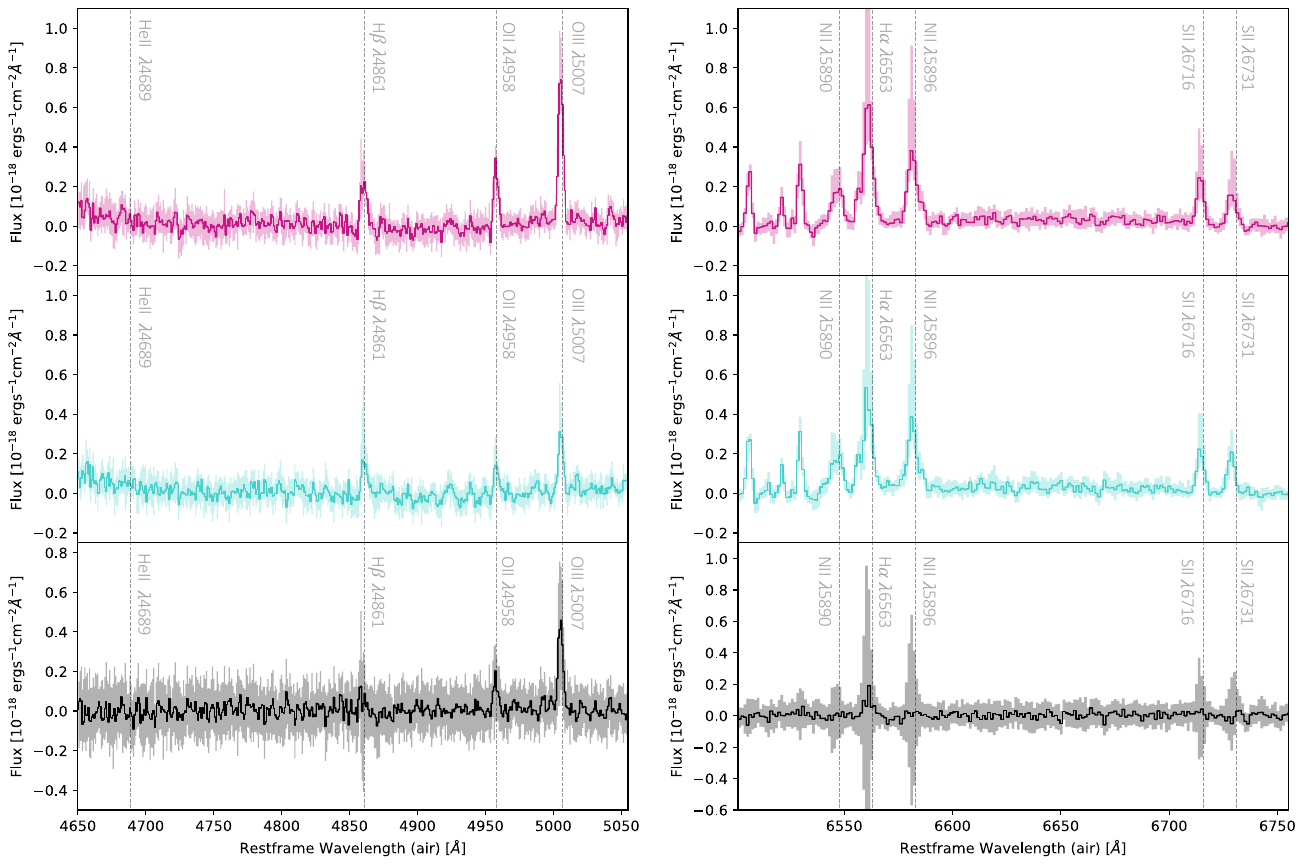}
        \caption{Average spectrum of the [O\,{\sc iii}] knots compared to nearby regions of the galaxy. \textit{Upper:} Mean spectrum for the [O\,{\sc iii}] knots (pink). \textit{Middle:} Average spectrum of nearby regions of the galaxy (blue). \textit{Lower:} Knot minus nearby region residual spectrum. Shaded regions denote the 32nd and 68th percentiles of the spectra in each case.}
    \label{fig:avgspec}
    \end{figure*}

    Once the positions of the [O\,{\sc iii}] knots have been identified, we extract their spectra, integrated within circular 0.4 arcsec (radius) apertures centred on each of the knots. Identical apertures are placed at a distance of 0.8 arcsec from each of the knots to provide reference spectra from nearby regions of the galaxy. We opt to select nearby regions rather than randomly placed apertures to eliminate any potential radial bias in the reference spectra. Each of the reference apertures is visually inspected to ensure it does not contain [O\,{\sc iii}] emission from any of the knots. The mean spectrum for the knots and reference apertures are shown in Fig.~\ref{fig:avgspec}. Based on the residual (knots minus nearby regions) spectrum, we find the knots have stronger [O\,{\sc iii}] emission than the underlying galaxy on average, with a peak [O\,{\sc iii}] flux $\sim$60 per cent brighter than that of the average reference spectrum. Conversely, the peak of both the [N\,{\sc ii}] and H$\alpha$ emission are weaker in the average knot spectrum, appearing $\sim$20-30 per cent stronger in the nearby regions of the galaxy. Here we discuss the potential ionising source(s) of these [O\,{\sc iii}] knots, presenting four possible scenarios: stellar clusters, AGN photoionisation, extended HII regions, and planetary nebulae (PNe).
    
        %...................................................................
        \subsubsection{Stellar clusters}
        \label{sec:stellar_clusters}
        %...................................................................
        
        One potential source of the [O\,{\sc iii}] knots observed in NGC\,4418 is stellar clusters: regions of the galaxy ionised by young stars. If the knot structures we observe are ionised in this way, we expect to detect them at shorter optical wavelengths where the light from young stars dominates. However, U (F336W) and V (F555W) band imaging from the HST (Gallagher et al. \textit{in prep.}) reveals no evidence for young stellar populations at the locations of the [O\,{\sc iii}] knots. Although two knots (K08 and K50) lie within $<$0.3 arcsec of detected HST sources, K50 corresponds to an unresolved pair of redder stellar sources, likely globular clusters, and K08 is detected only in the V band, suggesting it is an extremely bright PN. For the majority of the knots (49 of 51) we find no HST counterpart, meaning stellar clusters are likely not the dominant source of the [O\,{\sc iii}] knots in NGC\,4418. Furthermore, stellar clusters tend to follow the rotation of the galaxy (stellar) disk, yet the moment-1 velocity map of [O\,{\sc iii}] (Fig.~\ref{fig:ionised}) reveals that the knots are kinematically distinct from the galaxy disk. Based both on this kinematic information from MUSE, and the lack of detection in HST imaging, we therefore conclude that the [O\,{\sc iii}] knots in NGC\,4418 are not ionised by stellar clusters.
        
        %...................................................................
        \subsubsection{AGN photoionisation}
        \label{sec:agn_photoionisation}
        %...................................................................

\begin{figure}
        \centering
            \includegraphics[trim= 220 70 220 90 ,clip,width=.5\textwidth]{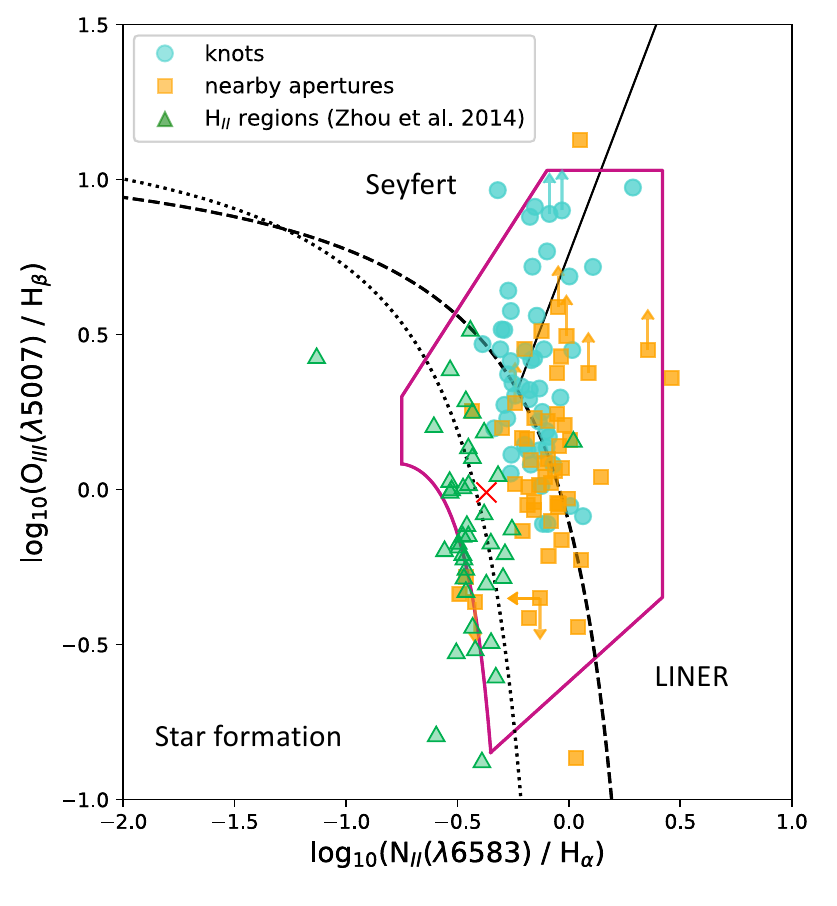}
            \caption{BPT diagram showing the dominant ionising source of the knots (blue circles) and nearby regions (yellow squares) compared to a sample of known HII regions in the nearby LIRG Arp 86 from \cite{zhou14} (green triangles), based on the diagnostic models of \cite{kauffmann03} (dotted line), \cite{kewley01} (dashed line), and \cite{kewley06} (solid line). The region of the diagram consistent with shock heating \citep{alatalo16} is overlaid (pink). Red cross marks the position of the central spaxel denoting the position of the CON in  NGC\,4418.}
        \label{fig:BPT}
        \end{figure}

        Another possible mechanism behind the knot structures seen in NGC\,4418 is the presence of so-called {AGN echoes}. These are regions of the galaxy that have been photoionised by an AGN and are predominantly characterised by their AGN-like line ratios on scales $>$10 kpc from the galaxy centre. To test whether AGN echoes are a plausible source of the [O\,{\sc iii}] knots observed in NGC\,4418, we measure the [O\,{\sc iii}]\,($\lambda$5007)\,/\,H$\beta$ and [N\,{\sc ii}]\,($\lambda$6583)\,/\,H$\alpha$ line ratios of each knot and plot them on a Baldwin, Phillips, and Terlevich (BPT) diagram (Fig.~\ref{fig:BPT}), designed to use line diagnostics to distinguish star formation from low-ionisation emission-line regions (LINERs) and Seyferts. In each case, spectra are integrated within a 0.4\,arcsec (radius) circular apertures, centred on the knot position and Gaussian profiles are fit to the relevant emission features. We highlight that all spectra are corrected for the dust extinction measured in section~\ref{sec:dust_extinction}, prior to the fitting. The fluxes for each line are then measured by integrating over the best-fit Gaussian line profiles. In cases where one of the features does not exceed the spectral noise, we instead measure an upper limit on the flux, assuming the same FWHM as the other features in the spectrum. These are shown as upper and lower limits on the line ratios. 

       Fig.~\ref{fig:BPT} shows all of the knots in NGC\,4418 to be inconsistent with heating via star formation alone. Even based on the more conservative constraints of \cite{kewley06}, 37 of the 51 knots (72.5 per cent) are classified as Seyfert- or LINER-like. The remaining 14 knots are consistent with a transitional phase. Objects in this transition region are often interpreted in the literature as either composite systems in which a central LINER is contaminated by surrounding  H\,{\sc ii} regions \citep[e.g.][]{ho93,menezes22}, or accretion powered systems with a lower accretion rate \citep[e.g.][]{ho08}. It has even been suggested that evolved hot stars and turbulent mixing layers in the interstellar medium (ISM) can also impact the H$\alpha$ emission in these transition objects \citep{shields07}. We further note that the LINER classification remains somewhat controversial, with low-ionisation AGN \citep[e.g.][]{ferland83,halpern83}, hot low-mass evolved stars (HOLMESs), or {post-asymptotic giant branch stars} \citep[e.g.][]{binette94,eracleous10,cid11} and shock heating \citep[e.g.][]{heckman80,dopita95,dopita96} all proposed to explain the emission line spectra of LINERs. Indeed, all but one of the knots and all but four of the nearby regions are also consistent with ionisation resulting from shock-heating \citep{alatalo16}. Nevertheless, the unanimous Seyfert- or LINER- like classification of the knots suggests the presence of an AGN in the CON of NGC\,4418.

        An AGN photoionisation scenario is further supported by the velocities of the ionised gas (Fig.~\ref{fig:ionised}), where we find the [O\,{\sc iii}] knots are kinematically independent from the rotation of the galaxy disk and are therefore likely to be part of a 3D gas structure surrounding the central disk. We note that while the knots appear to have been ionised by an AGN, albeit with a low ionisation parameter as indicated by the LINER classification, it is unlikely that the nucleus of NGC\,4418 is entirely powered by this AGN. Previous high resolution radio imaging of the nuclear region of NGC\,4418 reveals several compact features, thought to be super star clusters \citep{varenius14}. It is therefore likely that some additional star formation component is contributing to the central luminosity. This conclusion is consistent with our finding that several of the knots exist in the transition region of the diagram. Furthermore, we find the central spaxel, marking the position of the CON in NGC\,4418, to also lie in this transition region (red cross, Fig.~\ref{fig:BPT}). While the extreme dust obscuration in the nucleus, along with the spatial resolution of the MUSE imaging, severely limit our ability to interpret the emission line ratios of the CON itself, existing in the transition region is again consistent with the presence of an AGN with some starburst contribution. We therefore conclude that the CON in NGC\,4418 is likely powered by a combination of AGN and starburst activity.

        %...................................................................
        \subsubsection{Extended HII regions}
        \label{sec:extended_hii_regions}
        %...................................................................

        H\,{\sc ii} regions, or diffuse nebulae, are regions of interstellar gas that have been ionised by one or more O- or early B-type star(s). The spectra of H\,{\sc ii} regions typically contain strong hydrogen recombination lines and collisionally excited [N\,{\sc ii}] and [OIII], with additional [O\,{\sc iii}] and [N\,{\sc ii}] emission that varies strongly with the temperature of the ionising source. Due to their association with young stars, H\,{\sc ii} regions, like stellar clusters, are expected to be associated with optical U- and V- band emission. The lack of detection in the HST U (F336W) and V (F555W) band imaging (Gallagher et al. \textit{in prep.}), therefore, makes a young stellar source of ionisation unlikely. To further test this possibility, we compare the BPT line ratios of the knots in NGC\,4418 (Fig.~\ref{fig:BPT}) with a sample of 69 known H\,{\sc ii} regions in the nearby ($z\sim$0.0162), interacting system Arp86 \citep{zhou14}. Much like NGC\,4418, Arp 86 is a LIRG and is undergoing an interaction with a much smaller companion galaxy. These H\,{\sc ii} regions occupy a similar region in the BPT diagram to the reference spectra taken from regions of the galaxy close to the knots (Fig.~\ref{fig:BPT}). In general, the H\,{\sc ii} regions lie below the knots in the BPT diagram, with the majority of H\,{\sc ii} regions lying below the upper limit of star formation. Indeed, \cite{zhou14} conclude star formation to be the dominant ionising source related to the H\,{\sc ii} regions in Arp86, with a handful of regions consistent with both ionisation by star formation and the presence of shocks, AGB stars or AGN photoionisation. Nevertheless, Fig.~\ref{fig:BPT} shows a clear distinction between these H\,{\sc ii} regions and the knots identified in NGC\,4418, implying a different source of ionisation.
        
        %...................................................................
        \subsubsection{Planetary nebulae}
        \label{sec:planetary_nebulae}
        %...................................................................

        Finally, we considered the possibility that the [O\,{\sc iii}] knots trace PNe in NGC\,4418. PNe are shells of ionised gas that have been lost by stars. The remnant central stars of PNe are typically much older and less luminous than the O-type stars associated with H\,{\sc ii} regions and are thus not expected to be detected in HST imaging. One clear identifier of PNe is the presence of a helium recombination line, He\,{\sc ii} ($\lambda$4686), whose intensity exceeds a few per cent of H$\beta$. We do not detect this feature in the individual knot spectra or in the average spectrum of the knots (Fig.~\ref{fig:avgspec}), although we emphasise that the lack of this feature cannot rule out the existence of PNe. PNe can also be identified through an [O\,{\sc iii}]/H$\beta$ ratio $\gtrsim$4, which is {only} present in PNe. Indeed, we find the [O\,{\sc iii}]/H$\beta$ ratio in the knots generally exceeds that measured from the reference spectra. However, while six of the 51 knots ($\sim$12 per cent) have [O\,{\sc iii}]/H$\beta$ ratios $\gtrsim$4, the majority of the knots lie below this threshold and thus cannot be confirmed as PNe. Despite finding little evidence to support the PNe scenario, we highlight that these two diagnostics (the presence of He\,{\sc ii} and the [O\,{\sc iii}]/H$\beta$ line ratio) are typically biased against central stars with low effective temperatures, T$_{\rm{eff}}$, such as young PNe. We therefore cannot rule out young PNe as the source of the ionising radiation based on these diagnostics alone. We note however that PNe shells expand with velocities of a few tens of km\,s$^{-1}$, meaning they become rapidly unobservable on timescales of about a few 10$^4$ years. Given their short lifetimes, it is unlikely that we are seeing such widespread PNe simultaneously on scales $>$10 kpc in NGC\,4418. Additionally, at the redshift of NGC\,4418, PNe are expected to appear as point sources, yet we find many of the [OIII] knots to be extended and irregular. Finally, PNe are expected to follow the rotation of the galaxy disk, yet velocity information of the ionised gas shows this is not the case. We therefore suggest that PNe are unlikely to be the dominant ionising source of the [O\,{\sc iii}] knots in NGC\,4418, although they could be responsible for a small fraction of the knots observed. For the majority of the knots, we instead conclude AGN photoionisation to be the most likely cause.

%***************************************************************************
\section{Conclusions}
\label{sec:conclusions}
%***************************************************************************

We have presented new, targeted MUSE IFU observations for the local CON NGC\,4418. Our main conclusions are as follows;

    (i) For the first time, we have mapped the stellar velocity field of NGC\,4418. We find the stellar disk to be approaching on the north-eastern and receding on the south-western side with line-of-sight velocities of $V\sim$-115\,km\,s$^{-1}$ and $V\sim$80\,km\,s$^{-1}$, respectively. Within the central kiloparsec of the galaxy, we also reveal significant non-circular motions. We suggest these to be due to either a counter-rotating central stellar disk, perhaps caused by a recent interaction event, or the presence of a nuclear bar. While the data presented in this work are insufficient to conclude on the nature of this feature, we hope that future, high-resolution optical IFU studies will allow us to distinguish between these scenarios. 

    (ii) We confirm the presence of a kiloparsec-scale blueshifted outflow along the minor axis of NGC\,4418, extending in the north-west direction. The presence of this outflow has previously been postulated in the work of \cite{sakamoto13} and \cite{ohyama19}, where it is referred to as a dusty superwind. Indeed, we show this outflow to be associated with a region of enhanced velocity dispersion, and to be surrounded by a `U'-shaped dust structure. Our findings are consistent either with an outflow cone in which the edges appear dustier due to the cone's orientation or with a collimated outflow that has cleared the bulk of the dust from its path. Based on the line wings of H$\alpha$, we find the outflow to be decelerating, with a projected velocity of between -300$< V <$-140 km\,s$^{-1}$ at the outer tip. For the first time, we discover that this outflow also extends to the south-east, appearing blueshifted on both sides of the galaxy nucleus.    
    
    (iii) In addition to discovering the south-eastern side of the minor axis outflow, we identify several other new features of NGC\,4418. We detect a redshifted feature to the south composed of two seemingly distinct regions, S1 and S2. We conclude that this feature is likely outflowing material located behind the galaxy disk with an estimated velocity of 85$< V <$140 km\,s$^{-1}$, based on the H$\alpha$ line wings. This redshifted outflow appears to be connected to a `tail' of high velocity gas surrounding the galaxy, which we refer to as the southern tail. In addition to these redshifted features, we also discover a bubble-like structure to the north of the galaxy nucleus: the northern bubble. The highest velocity gas ($V <$-140 km\,s$^{-1}$) traces the edges of the bubble, whilst the apparent void of the bubble consists of slower gas ($V >$-85 km\,s$^{-1}$) with enhanced H$\beta$ absorption and dust extinction. We propose that the observed bubble could either be an accelerating shell of gas expanding outwards from the galaxy centre or part of the minor axis outflow structure beginning to infall back onto the plane of the galaxy and appearing as a bubble due to orientation effects.

    (iv) For the first time, we uncover the presence of [O\,{\sc iii}] knots throughout NGC\,4418. Their lack of detection in the HST U (F336W) and V (F555W) bands, AGN-like line ratios, and distinct velocity profiles imply they are likely regions of the galaxy that have been photoionised by an AGN. From this, we postulate that the CON in NGC\,4418 is powered by AGN activity, with the photoionised knots tracing the dust structure of the inner nuclear region. 

Overall, our results are consistent with NGC\,4418 being in a phase in the AGN feedback cycle where dust is being expelled from the galactic nucleus via outflows. The large-scale structure of NGC\,4418 shows evidence for regions ionised by an AGN. Along with the fast outflows ($V >$140 km\,s$^{-1}$) and bubble structures, we suggest it is unlikely that the CON is powered solely by starburst activity. Instead, we suggest NGC\,4418 has either recently undergone or is currently in an active AGN phase. Future IFU studies of the wider CON population are needed to determine whether all CONs are consistent with this classification. If so, the number of AGN in the local Universe may double with respect to what is currently estimated, completely revolutionising our understanding of how AGN form and their role in galaxy evolution as a whole.

%***************************************************************************
% Acknowledgements
%***************************************************************************

\begin{acknowledgements}
      C.W. gratefully acknowledges support from a European Research Council (ERC) Advanced Grant, 789410. The work of YN was supported by JSPS KAKENHI grant No. JP23K13140.
\end{acknowledgements}

%***************************************************************************
% REFERENCES
%***************************************************************************
\bibliographystyle{aa}
\bibliography{refs}

%***************************************************************************
% Appendices
%***************************************************************************

\begin{appendix} 
%First appendix
\counterwithin{figure}{section}

\section{Supplementary figures and tables}

\begin{figure*}[htp]
\centering
\includegraphics[trim= 150 150 170 150,clip,width=\textwidth]{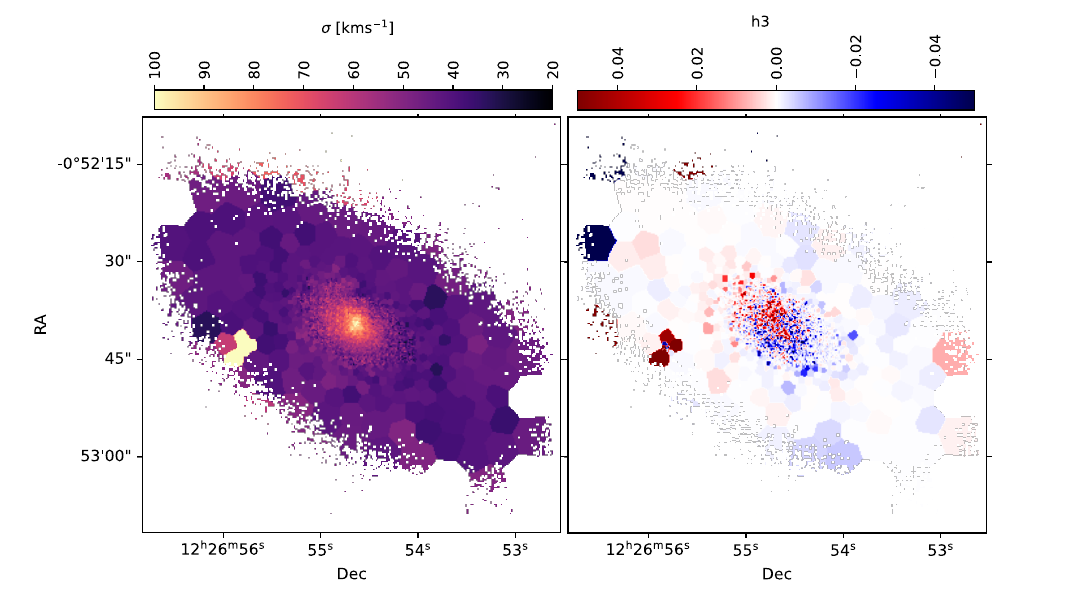}
            \caption{Stellar kinematics of NGC\,4418 output from the pPXF fitting. \textit{Left:} Map of the stellar velocity dispersion, $\sigma$. \textit{Right:} Map of the h3 Gauss-Hermite coefficient. In each case, pixels have been binned to achieve a signal-to-noise ratio of 20. An initial signal-to-noise ratio cut of 2 has been applied, with pixels below this threshold omitted from the binning.}
    \label{fig:veldisp_h3}
\end{figure*}

\begin{table*}
    \centering
    \begin{tabular}{c|c|c|c|c}
    \hline
    Knot ID    &  RA (deg) &  Dec (deg)    &  RA (J200) &  Dec (J200)  \\
    \hline
    K01 & 186.7285 & -0.8842 & 12:26:54.8319 & -0:53:02.9919 \\
    K02 & 186.7249 & -0.8839 & 12:26:53.9797 & -0:53:02.0972 \\
    K03 & 186.7228 & -0.8827 & 12:26:53.4798 & -0:52:57.6230 \\
    K04 & 186.7271 & -0.8824 & 12:26:54.5129 & -0:52:56.7076 \\
    K05 & 186.7213 & -0.8823 & 12:26:53.1209 & -0:52:56.1114 \\
    K06 & 186.7261 & -0.8814 & 12:26:54.2610 & -0:52:52.8641 \\
    K07 & 186.7236 & -0.8814 & 12:26:53.6609 & -0:52:53.0114 \\
    K08 & 186.7292 & -0.8813 & 12:26:55.0003 & -0:52:52.5563 \\
    K09 & 186.7233 & -0.8810 & 12:26:53.5869 & -0:52:51.4411 \\
    K10 & 186.7298 & -0.8804 & 12:26:55.1441 & -0:52:49.4525 \\
    K11 & 186.7225 & -0.8802 & 12:26:53.4036 & -0:52:48.6314 \\
    K12 & 186.7270 & -0.8800 & 12:26:54.4727 & -0:52:47.8284 \\
    K13 & 186.7259 & -0.8798 & 12:26:54.2130 & -0:52:47.1355 \\
    K14 & 186.7325 & -0.8788 & 12:26:55.8074 & -0:52:43.5477 \\
    K15 & 186.7313 & -0.8799 & 12:26:55.5145 & -0:52:47.6114 \\
    K16 & 186.7280 & -0.8798 & 12:26:54.7197 & -0:52:47.1688 \\
    K17 & 186.7209 & -0.8797 & 12:26:53.0117 & -0:52:46.9480 \\
    K18 & 186.7251 & -0.8790 & 12:26:54.0210 & -0:52:44.2481 \\
    K19 & 186.7236 & -0.8791 & 12:26:53.6609 & -0:52:44.6114 \\
    K20 & 186.7325 & -0.8787 & 12:26:55.8092 & -0:52:43.2477 \\
    K21 & 186.7267 & -0.8789 & 12:26:54.4143 & -0:52:43.9114 \\
    K22 & 186.7281 & -0.8787 & 12:26:54.7531 & -0:52:43.4307 \\
    K23 & 186.7225 & -0.8782 & 12:26:53.4054 & -0:52:41.4955 \\
    K24 & 186.7268 & -0.8781 & 12:26:54.4270 & -0:52:41.2389 \\
    K25 & 186.7281 & -0.8779 & 12:26:54.7377 & -0:52:40.3342 \\
    K26 & 186.7275 & -0.8780 & 12:26:54.6077 & -0:52:40.8114 \\
    K27 & 186.7250 & -0.8779 & 12:26:54.0020 & -0:52:40.3366 \\
    K28 & 186.7274 & -0.8775 & 12:26:54.5852 & -0:52:38.9465 \\
    K29 & 186.7290 & -0.8769 & 12:26:54.9543 & -0:52:36.8135 \\
    K30 & 186.7261 & -0.8777 & 12:26:54.2529 & -0:52:39.5539 \\
    K31 & 186.7234 & -0.8775 & 12:26:53.6220 & -0:52:39.1011 \\
    K32 & 186.7270 & -0.8770 & 12:26:54.4848 & -0:52:37.1241 \\
    K33 & 186.7304 & -0.8765 & 12:26:55.2969 & -0:52:35.4750 \\
    K34 & 186.7337 & -0.8764 & 12:26:56.0970 & -0:52:35.1167 \\
    K35 & 186.7326 & -0.8763 & 12:26:55.8327 & -0:52:34.7045 \\
    K36 & 186.7316 & -0.8761 & 12:26:55.5844 & -0:52:34.0764 \\
    K37 & 186.7264 & -0.8759 & 12:26:54.3354 & -0:52:33.2953 \\
    K38 & 186.7280 & -0.8759 & 12:26:54.7265 & -0:52:33.3392 \\
    K39 & 186.7301 & -0.8757 & 12:26:55.2193 & -0:52:32.5183 \\
    K40 & 186.7335 & -0.8756 & 12:26:56.0412 & -0:52:32.3114 \\
    K41 & 186.7327 & -0.8755 & 12:26:55.8489 & -0:52:31.8991 \\
    K42 & 186.7257 & -0.8754 & 12:26:54.1610 & -0:52:31.6114 \\
    K43 & 186.7343 & -0.8751 & 12:26:56.2342 & -0:52:30.2879 \\
    K44 & 186.7273 & -0.8747 & 12:26:54.5453 & -0:52:28.7759 \\
    K45 & 186.7341 & -0.8743 & 12:26:56.1946 & -0:52:27.4114 \\
    K46 & 186.7276 & -0.8738 & 12:26:54.6319 & -0:52:25.5774 \\
    K47 & 186.7311 & -0.8738 & 12:26:55.4572 & -0:52:25.6667 \\
    K48 & 186.7335 & -0.8738 & 12:26:56.0345 & -0:52:25.7638 \\
    K49 & 186.7302 & -0.8734 & 12:26:55.2493 & -0:52:24.2681 \\
    K50 & 186.7266 & -0.8733 & 12:26:54.3769 & -0:52:23.9732 \\
    K51 & 186.7246 & -0.8731 & 12:26:53.8921 & -0:52:23.1144 \\
    K52 & 186.7328 & -0.8725 & 12:26:55.8812 & -0:52:21.0114 \\
    K53 & 186.7297 & -0.8713 & 12:26:55.1270 & -0:52:16.6783 \\
    \hline
    \end{tabular}
    \caption{World coordinate system (WCS) positions of the [O\,{\sc iii}] $\lambda5007$ knots identified in NGC\,4418.}
    \label{tab:a1}
\end{table*}
\end{appendix}

%%%% End of aa.dem
\end{document}